\documentstyle[12pt]{article}

\def\hybrid{\topmargin -20pt  \oddsidemargin 0pt
      \headheight 0pt   \headsep 0pt
      \textwidth 6.25in 
      \textheight 9.5in 
      \marginparwidth .875in
      \parskip 5pt plus 1pt   \jot = 1.5ex}

\hybrid

\begin{document}
\def\x{\times}
\def\beq{\begin{equation}}
\def\eeq{\end{equation}}
\def\beqa{\begin{eqnarray}}
\def\eeqa{\end{eqnarray}}
\def\D{ {\cal D}}
\def\L{ {\cal L}}
\def\C{ {\cal C}}
\def\N{ {\cal N}}
\def\calE{{\cal E}}
\def\lin{{\rm lin}}
\def\Tr{{\rm Tr}}
\def\mxth{\mathsurround=0pt }
\def\xversim#1#2{\lower2.pt\vbox{\baselineskip0pt \lineskip-.5pt
x  \ialign{$\mxth#1\hfil##\hfil$\crcr#2\crcr\sim\crcr}}}
\def\simgr{\mathrel{\mathpalette\xversim >}}
\def\simle{\mathrel{\mathpalette\xversim <}}

\def\a{\alpha}
\def\b{\beta}
\def\dota{ {\dot{\alpha}} }
\def\lag{Lagrangian}
\def\Kahler{K\"{a}hler}
\def\kahler{K\"{a}hler}
\def\A{ {\cal A}}
\def\C{ {\cal C}}
\def\D{ {\cal D}}
\def\F{{\cal F}}
\def\L{ {\cal L}}

\def\R{ {\cal R}}
\def\x{ \times }
\def\beq{\begin{equation}}
\def\eeq{\end{equation}}
\def\beqa{\begin{eqnarray}}
\def\eeqa{\end{eqnarray}}

\sloppy
\newcommand{\be}{\begin{equation}}
\newcommand{\eq}{\end{equation}}
\newcommand{\ov}{\overline}
\newcommand{\un}{\underline}
\newcommand{\p}{\partial}
\newcommand{\la}{\langle}
\newcommand{\ra}{\rangle}
\newcommand{\bl}{\boldmath}
\newcommand{\ds}{\displaystyle}
\newcommand{\nl}{\newline}
\newcommand{\Nzahl}{{\bf N}  }
\newcommand{\zzahl}{ {\bf Z} }
\newcommand{\Zzahl}{ {\bf Z} }
\newcommand{\Qzahl}{ {\bf Q}  }
\newcommand{\Rzahl}{ {\bf R} }
\newcommand{\Czahl}{ {\bf C} }
\newcommand{\wt}{\widetilde}
\newcommand{\wh}{\widehat}
\newcommand{\fs}[1]{\mbox{\scriptsize \bf #1}}
\newcommand{\ft}[1]{\mbox{\tiny \bf #1}}
\newtheorem{satz}{Satz}[section]
\newenvironment{Satz}{\begin{satz} \sf}{\end{satz}}
\newtheorem{definition}{Definition}[section]
\newenvironment{Definition}{\begin{definition} \rm}{\end{definition}}
\newtheorem{bem}{Bemerkung}
\newenvironment{Bem}{\begin{bem} \rm}{\end{bem}}
\newtheorem{bsp}{Beispiel}
\newenvironment{Bsp}{\begin{bsp} \rm}{\end{bsp}}
\renewcommand{\arraystretch}{1.5}



\renewcommand{\thesection}{\arabic{section}}
\renewcommand{\theequation}{\thesection.\arabic{equation}}

\parindent0em

\def\S4{\frac{SO(4,2)}{SO(4) \otimes SO(2)}}
\def\P3{\frac{SO(3,2)}{SO(3) \otimes SO(2)}}
\def\MGd{\frac{SO(r,p)}{SO(r) \otimes SO(p)}}
\def\SOd{\frac{SO(r,2)}{SO(r) \otimes SO(2)}}
\def\SO2{\frac{SO(2,2)}{SO(2) \otimes SO(2)}}
\def\SUm{\frac{SU(n,m)}{SU(n) \otimes SU(m) \otimes U(1)}}
\def\SUS{\frac{SU(n,1)}{SU(n) \otimes U(1)}}
\def\SK{\frac{SU(2,1)}{SU(2) \otimes U(1)}}
\def\SU{\frac{ SU(1,1)}{U(1)}}

\begin{titlepage}
\begin{center}

\hfill HUB-IEP-95/12\\
\hfill DESY 95-??\\
\hfill hep-th/9507113\\

\vskip .1in

{\bf NON-PERTURBATIVE MONODROMIES IN $N=2$ HETEROTIC STRING VACUA}

\vskip .2in

{\bf Gabriel Lopes Cardoso,
Dieter L\"{u}st }  \\
\vskip .1in

{\em  Humboldt Universit\"at zu Berlin\\
Institut f\"ur Physik\\
D-10115 Berlin, Germany}\footnote{e-mail addresses:
GCARDOSO@QFT2.PHYSIK.HU-BERLIN.DE, LUEST@QFT1.PHYSIK.HU-BERLIN.DE}

\vskip .1in
and
\vskip .1in

{\bf Thomas Mohaupt}
\vskip .1in

{\em
DESY-IfH Zeuthen\\
Platanenallee 6\\
D-15738 Zeuthen, Germany}\footnote{e-mail address: MOHAUPT@HADES.IFH.DE}

\end{center}

\vskip .2in

\begin{center} {\bf ABSTRACT } \end{center}
\begin{quotation}\noindent
We address non-perturbative effects and duality symmetries in $N=2$
heterotic string theories in four dimensions.  Specifically,
we consider how each of the
four lines of enhanced gauge symmetries in the perturbative
moduli space of $N=2$ $T_2$ compactifications is split into 2 lines where
monopoles and dyons become massless.  This amounts to considering
non-perturbative effects originating from enhanced gauge symmetries
at the microscopic string level.
We show that the perturbative
and non-perturbative monodromies consistently lead to the results
of Seiberg-Witten upon identication of a consistent truncation
procedure from local to rigid $N=2$ supersymmetry.
\end{quotation}
July 1995\\
\end{titlepage}
\vfill
\eject

\newpage

\section{Introduction}

Recently some major progress
has been obtained in the understanding  of the non-perturbative
dynamics of $N=2$ Yang-Mills theories as well as $N=2$ superstrings
in four dimensions.
The quantum moduli space of the $N=2$ $SU(r+1)$
Yang-Mills theories is described \cite{SW1,SW2,KLTY,KLT}
by {\it rigid special geometry}, which is based on an (auxiliary)
Riemann surface of genus $r$ with the $2r$ periods $(a_i,a_{Di})$
$(r=1,\dots r)$ being
holomorphic sections of the $Sp(2r,{\bf Z})$ bundle defined
by the first
homology group of the Riemann surface. The periods $a_i$ are associated
to $r$ $N=2$ vector superfields in the Cartan subalgebra of $SU(r+1)$;
hence non-zero vacuum expectation values break $SU(r+1)$ down
to $U(1)^{r+1}$. The quantum moduli space possesses singular points with
non-trivial monodromies around these points. The semiclassical
monodromies are due to the one-loop contributions to the
holomorphic prepotential, and the corresponding logarithmic singularities
are the left-over signal of the additional
non-Abelian massless vector fields at $a_i=0$.
However in the full quantum moduli space there are no points of enhanced
non-Abelian gauge symmetries, and the semiclassical monodromies are
split into non-perturbative monodromies,
where the monodromy group $\Gamma_M$ is a subgroup of $Sp(2r,{\bf Z})$.
The corresponding singular
points in the quantum moduli space are due to magnetic monopoles and
dyons becoming massless at these points.

During the past years duality symmetries in string theory gained a lot
of attention. First, the socalled target space duality symmetry ($T$-duality)
(for a review see \cite{GPR}) is
known to be a true symmetry in every order of
string perturbation theory. Second, $S$-duality \cite{FILQ} was proposed to
be a non-perturbative string symmetry, and evidence for $S$-duality in
$N=4$ heterotic strings is now accumulating. Moreover,
string-string dualities
\cite{many1,KacVaf} between  type II,I and heterotic theories  in various
different dimensions play an important role in the understanding of string
dynamics.

In this paper we will address non-perturbative effects
and duality symmetries in $N=2$ heterotic string
theories in four dimensions.
When coupling the $N=2$ Yang-Mills gauge theory
to supergravity, as it is necessary in the
context of superstrings, different additional effects play an important role.
First, there is always one additional $U(1)$ vector field, the socalled
graviphoton. As a result of the gravitational interactions the
couplings of $n$
$N=2$ vector multiplets plus
the graviphoton are now described by {\it local special  geometry} with
the
$2n+2$ holomorphic periods $(X^I,iF_I)$ ($I=0,\dots ,n)$ being
holomorphic
sections of an $Sp(2n+2)$ bundle \cite{many2}. Due
to the absence of a propagating
scalar partner of the graviphoton, the associated special K\"ahler space
can be parametrized by projective, {\it special} coordinates $z^A=X^A/X^0$
($A=1,\dots ,n)$. In the context of  four-dimensional $N=2$ heterotic
string vacua another $U(1)$ gauge boson plus a corresponding (complex)
scalar field
exists besides the vector
fields of the rigid theory, namely the dilaton-axion field $S$.
In the rigid theories, $S$ appears merely as a parameter for
the classical gauge coupling plus theta angle, but in string theories $S$
becomes a dynamical field. It can be either described by a vector-tensor
multiplet \cite{SohSteWes,WKLL}
or, as we will keep it in the following, by an $N=2$
vector multiplet. Thus the total number of physical vector multiplets
is given by $n=r+1$, where $r$ is the number of moduli fields
in vector multiplets which are in one to one correspondence
with the Higgs fields of the rigid theory. In a recent very interesting
developement, some convincing evidence accumulated
\cite{KacVaf,many3} that the periods
$(X^I,iF_I)$ of the full heterotic $N=2$ quantum theory are given
by a suitably chosen Calabi-Yau threefold with dimension
of the third cohomology group $b_3=4+2r$. Moreover, based on the ideas
of heterotic versus type II string duality, this Calabi-Yau space
does not just serve as an auxiliary construction, but there exists a
dual type II, $N=2$ string compactified on this Calabi-Yau space.
This observation opens the exiting possibility to obtain non-perturbative
quantum effects on the heterotic side by computing the classical vector
couplings on the type II side, since in the type II theories the
dilaton as the loop counting parameter sits in a hyper multiplet \cite{Sei}
and, at lowest order, does not couple to the type II vector fields.
If this picture turns out to be true, it consequently implies
that the Riemann surface of the rigid theory is embedded into
the six-dimensional Calabi-Yau space.

In this paper we investigate
(however without reference to an underlying Calabi-Yau
space) how the semiclassical singular lines
of enhanced gauge symmetries in  $N=2$ heterotic strings can be
split non-perturbatively each into two singular loci, namely each into two
lines
where magnetic monopoles or dyons respectively beome massless.
Specifically, under the (reasonable) assumption that the non-perturbative
dynamics well below the Planck scale is governed by the Yang-Mills gauge
interactions a la Seiberg and Witten \cite{SW1}, we are able to construct
the associated non-perturbative monopol and dyon
monodromy matrices.
In addition, we address the question,  how the already known
perturbative as well as the here newly derived
non-perturbative monodromies of the $N=2$ heterotic moduli space
lead to the rigid monodromies of \cite{SW1,KLTY,KLT}. This embedding of the
rigid monodromies into the local ones implies a very well defined
truncation procedure. As we will show this does not agree with
the naive limit of $M_{\rm Planck}\rightarrow\infty$, because
one also has to take into account the fact that in the string case the
dilaton as well as the graviphoton are in general not invariant under
the Weyl transformation. Thus, in other words, the dilaton and graviphoton
fields have to be frozen, before one can perform the limit $M_{\rm Planck}
\rightarrow\infty$.

The classical as well as the perturbative (one-loop) holomorphic
prepotentials for $N=2$ heterotic strings were derived in
\cite{WKLL,CAFP,AFGNT}.
In particular, \cite{WKLL,AFGNT} focused on heterotic string vacua
which are given as a compactification of the six-dimensional $N=1$
heterotic string on a two-dimensional torus  $T_2$. These types of
backgrounds always lead to two moduli fields, $T$ and $U$, of
the underlying $T_2$; the underlying holomorphic prepotential is then
a function of $S$, $T$ and $U$. At special lines (points) in the
perturbative $(T,U)$ moduli space, part of the Abelian gauge group is
enhanced to $SU(2)$, $SU(2)^2$ or $SU(3)$ respectively. This is the
stringy version of the Higgs effect with Higgs fields given as certain
combinations of the moduli $T$ and $U$. Thus this situation is completely
analogeous to the rigid case discussed in \cite{SW1,SW2,KLTY,KLT};
in the string case,
however, the
Weyl group, acting on the Higgs fields $a_1$ and $a_2$,
 of $SU(2)$, $SU(2)^2$ or $SU(3)$ as the classical symmetry
group of the effective action is extended to be the full target space
duality group acting on the moduli fields $T$ and $U$
\cite{DVV}. In the next chapter
we will recall the classical prepotential and the classical duality
symmetries; in
particular we will work out
the precise relation between the field theory Higgs
fields and the string moduli, and
the relation of the four indepedent Weyl transformations to
specific elements of the target space duality group.
At the one loop level, the holomorphic prepotential exhibits logarithmic
singularities precisely at the critical lines of enhanced gauge
symmetries;
moving in moduli space around the critical  lines one obtaines
the semiclassical monodromies. In the third chapter we will
determine the $Sp(8,{\bf Z})$ one-loop monodromy matrices
corresponding to the four independent Weyl transformations
of the enhanced gauge groups. We will discuss the consistent truncation
to the rigid case and show that the truncated one-loop monodomies agree
with the semi-classical monodromies obtained in \cite{SW1,KLTY,KLT}.
Finally, in chapter four, we derive, under a few
physical assumptions, the splitting of
the one-loop (i.e. semi-classical) monodromies into a pair of non-perturbative
monopole and dyon monodromies. We show that with the same truncation
procedure as in the one loop case we arrive at the non-perturbative monodromies
of the rigid cases.  This analyis adresses the non-perturbative effects
in the gauge sectors far below the Planck scale. Thus the
dilaton field is kept large at the
points where monopoles or dyons become massless. In addition one
expects \cite{KacVaf} non-perturbative, genuine
stringy monodromies at finite values of the
dilaton, e.g. when gravitational instantons, black
holes etc. become massless. It would be of course
very interesting to compare our results with some computations
of monodromies in appropriately choosen
(such as $X_{24}(1,1,2,8,12)$ \cite{KacVaf}) Calabi-Yau moduli spaces.

\section{Classical results, enhanced gauge symmetries and Weyl
reflections \label{classres}}

In this section we collect some results
about $N=2$ heterotic strings and the related classical prepotential;
we will in
particular work out the
relation between the enhanced gauge symmtries, the duality
symmetries and Weyl transformations. We will consider four-dimensional
heterotic vacua which are based on compactifications of
six-dimensional vacua on a two-torus $T_2$. The moduli of $T_2$
are commonly denoted by $T$ and $U$ where $U$ describes the
deformations of the complex structure, $U=(\sqrt G-iG_{12})/G_{11}$
($G_{ij}$ is the metric of $T_2$), while $T$ parametrizes the
deformations of the area and of the antisymmetric tensor, $T=2(\sqrt G+iB)$.
(Possibly other existing vector fields will not play any role in our
discussion.)
The scalar fields $T$ and $U$ are the spin-zero components of
two $U(1)$ $N=2$ vector supermultiplets.
All physical properties of the two-torus compactifications are
invariant under the group $SO(2,2,{\bf Z})$ of discrete target space
duality transformations. It contains the $T\leftrightarrow U$
exchange, with group element denoted by $\sigma$ and the
$PSL(2,{\bf Z})_T\times PSL(2,{\bf Z})_U$ dualities, which act on $T$
and $U$ as
\be
(T,U) \longrightarrow \left(
\frac{aT - ib}{icT + d}, \frac{a'U - ib'}{ic'U +d'} \right),\;\;\;
a,b,c,d,a',b', c',d' \in {\bf Z},\;\;\;ad-bc = a'd' -b'c' =1.
\eq
The classical monodromy group,
which is a true symmetry of the classical effective Lagrangian,
is generated by the elements $\sigma$,
$g_1$, $g_1$: $T\rightarrow 1/T$ and $g_2$, $g_2$: $T\rightarrow 1/(T-i)$.
The transformation $t$: $T\rightarrow
T+i$, which is of infinite order, corresponds to $t=g_2^{-1}g_1$.
Whereas $PSL(2,{\bf Z})_T$ is generated by $g_1$ and $g_2$,
the corresponding elements in $PSL(2,{\bf Z})_U$ are obtained
by conjugation with $\sigma$, i.e. $g_i'o=\sigma^{-1}g_i\sigma$.

As mentioned already in the introduction, the $N=2$ heterotic string vacua
contain two further $U(1)$ vector fields, namely the graviphoton field,
which has no physical scalar partner, and the dilaton-axion field,
denoted by $S$. Thus the full Abelian gauge symmetry we consider is
given by $U(1)_L^2\times U(1)_R^2$.
At special lines in the $(T,U)$ moduli space, additional vector fields
become massless and the $U(1)_L^2$ becomes enlarged to a
non-Abelian gauge symmetry. Specifically, there are four
inequivalent lines in the moduli space where two charged gauge bosons
become massless. The quantum numbers of the states that become massless
can be easily read of from the holomorphic mass formula \cite{OogVaf,FKLZ,CLM}
\begin{equation}
{\cal M}=m_2-im_1U+in_1T-n_2TU,
\label{HMF}
\end{equation}
where $n_i$, $m_i$ are the winding and momentum quantum numbers
associated with the $i$-th direction of the target space $T_2$.
Let us now collect the fixed lines, the quantum numbers of the
states which become massless at the fixed lines; we already include
in the following table the Weyl transformations under which the
corresponding lines are fixed:
\be
\begin{array}{|l|l|l|} \hline
\mbox{FP Transformations} & \mbox{Fixed Points}
& \mbox{Quantum Numbers} \\ \hline
w_1 & U=T & m_1 = n_1 = \pm 1,\;\;\;m_2 = n_2 = 0\\
w_2 = w_1' & U = \frac{1}{T} &
m_2 = n_2 = \pm 1,\;\;\;m_1 = n_1 = 0\\
w_2' & U=T-i & m_1 = m_2 =n_1 = \pm 1,\;\;\;
n_2 =0\\
w_0' & U = \frac{T}{iT +1}
& m_1 = n_1 = -n_2 = \pm 1,\;\;\;m_2 =0\\ \hline
\end{array}
\label{CriticalLines}
\eq
At each of the four critical lines the $U(1)_L^2$ is extended to
$SU(2)_L\times U(1)_L$. Moreover, these lines intersect one another in two
inequivalent critical points (for a detailed discussion see (\cite{CLM})).
At $(T,U)=(1,1)$ the first two lines
intersect. The four extra massless states extend the
gauge group to $SU(2)_L^2$. At $(T,U)=(\rho,\bar\rho)$ ($\rho=e^{i\pi/6}$)
the last three lines intersect. The six additional states extend the
gauge group to $SU(3)$. (In addition, the first and the third line
intersect at $(T,U)=(\infty,\infty)$, whereas the first and the last
line intersect at $(T,U) = (0,0)$.)

The Weyl groups of the enhanced gauge groups $SU(2)^2$ and $SU(3)$,
realized at $(T,U)=(1,1),(\rho,\bar\rho)$ respectively,
have the following action on $T$ and $U$:
\be
\begin{array}{|l|l|l|} \hline
\mbox{Weyl Reflections} & T \rightarrow T' & U \rightarrow U' \\ \hline
w_1 & T \rightarrow U & U \rightarrow T \\
w_2 & T \rightarrow \frac{1}{U} & U \rightarrow \frac{1}{T}\\
w_1'&  T \rightarrow \frac{1}{U} & U \rightarrow \frac{1}{T}\\
w_2'&  T \rightarrow U+i & U \rightarrow T-i \\
w_0'& T \rightarrow \frac{U}{-iU+1} & U \rightarrow
\frac{T}{iT+1} \\ \hline
\end{array}
\label{WeylGroup}
\eq
$w_1$, $w_2$ are the Weyl reflections of $SU(2)_{(1)}\times SU(2)_{(2)}$,
whereas $w'_1$ and $w'_2$ are the fundamental Weyl reflections of the
enhanced $SU(3)$. For later reference we have also listed the $SU(3)$
Weyl reflection $w'_0={w'_2}^{-1} w'_1 w'_2$ at the hyperplane
perpendicular to the highest root of $SU(3)$. Note that $w_2=w'_1$.
All these Weyl transformations are target space modular transformations
and therefore elements of the monodromy group. All Weyl reflections
can be expressed in terms of the generators $g_1,g_2,\sigma$ and,
moreover, all Weyl reflections are conjugated to the mirror
symmetry $\sigma$ by some group element:
\be
w_1 = \sigma,\;\;\;
w_2 = w_1' = g_1 \sigma g_1 = g_1^{-1} \sigma g_1
\eq
\be
w_2' = t \sigma t^{-1} = (g_1^{-1} g_2)^{-1} \sigma (g_1^{-1}
g_2),\;\;\;
w_0' = w_2'^{-1} w_1' w_2'
\label{ConjugatedToSigma}
\eq
As already mentioned the four critical lines are fixed under the corresponding
Weyl transformation. Thus it immediately follows that the numbers
of additional massless states agrees with the order of the fixed
point transformation at the critical line, points respectively \cite{CLM}.

Let us now express the moduli fields $T$ and $U$ in terms
of the field theory Higgs fields whose non-vanishing vacuum
expectation values spontaneously break the enlarged gauge
symmetries $SU(2)^2$, $SU(3)$ down to $U(1)^2$.
First, the Higgs field\footnote{Note that the
Higgs fields just correspond to the uniformizing variables
of modular functions at the critical points, lines respectively.}
 of $SU(2)_{(1)}$
is given by $a_1 \propto (T-U)$.
Taking the rigid field theory limit $\kappa^2={8\pi\over M_{\rm Planck}^2}
\rightarrow 0$ we will expand $T=T_0+\kappa \delta T$, $U=T_0+\kappa\delta U$.
Then, at the linearized level, the $SU(2)_{(1)}$ Higgs field is given
as $a_1  \propto (\delta T-\delta U) $.
Analogously, for the enhanced $SU(2)_{(2)}$ the Higgs field is $a_2
\propto (T-1/U)$.
Again, we expand as $T=T_0(1+\kappa\delta T)$, $U={1\over T_0}(1+\delta U)$
which leads to $a_2 \propto \delta T+\delta U$.
Finally, for the enhanced $SU(3)$ we obtain as Higgs fields
$a'_1 \propto \delta T+\delta U$,
$a_2' \propto \rho^2 \delta T + \rho^{-2} \delta U$,
where we have expanded as $T=\rho+\delta T$, $U=\rho^{-1}+\delta U$
(see section \ref{pertsec} for details).

The classical vector couplings are determined by the holomorphic prepotential
which is a homogeneous function of degree two of the fields $X^I$
($I=1,\dots , 3$). It is given by \cite{CAFP,WKLL,AFGNT}
\begin{equation}
F=i{X^1X^2X^3\over X^0}=- STU,
\label{classprep}
\end{equation}
where the physical vector fields are defined as $S=i{X^1\over X^0}$,
$T=-i{X^2\over X^0}$, $U=-i{X^3\over X^0}$ and the graviphoton corresponds
to $X^0$. As explained in \cite{CAFP,WKLL},
the period vector $(X^I,iF_I)$ ($F_I=
{\partial F\over X^I}$), that follows from the prepotential (\ref{classprep}),
does not lead to classical gauge couplings which all become small in the
limit of large $S$. Specifically, the gauge couplings which involve
the $U(1)_S$ gauge group are constant or even grow in the string weak
coupling limit $S\rightarrow\infty$. In order to choose a `physical'
period vector one has to replace $F_{\mu\nu}^S$ by its dual which is
weakly coupled in the large $S$ limit. This is achieved by the
following symplectic transformation $(X^I,iF_I)\rightarrow (P^I,iQ_I)$
where\footnote{Note however that the new coordinates $P^I$ are not
independent and hence there is no prepotential $Q(P^I)$ with the property
$Q_I={\partial Q\over\partial P^I}$.}
\be
P^1 = i F_1,\; Q_1 = i X^1,\mbox{   and   }P^i = X^i,\;Q_i = F_i\quad
{\mbox {\rm for } }\quad i =0,2,3.
\label{PQtoXF}
\eq
In this new basis the classical period vector takes the form
\begin{equation}
\Omega^T=(1,TU,iT,iU,iSTU,iS,-SU,-ST),
\end{equation}
where $X^0=1$. One sees that after the transformation (\ref{PQtoXF})
all electric vector fields $P^I$ depend only on $T$ and $U$, whereas
the magnetic fields $Q_I$ are all proportional to $S$.

The basis $\Omega$ is also well adapted to discuss the action
of the target space duality transformations and, as particular
elements of the target space duality group, of the four inequivalent
Weyl reflections given in (\ref{WeylGroup}).
In general, the field equations of
the $N=2$ supergravity action are invariant under the following
symplectic $Sp(8,{\bf Z})$ transformations, which act on the period
vector $\Omega$ as
\begin{equation}
\pmatrix{P^I\cr i Q_I\cr}\rightarrow \Gamma\pmatrix{
P^I\cr  i Q_I\cr}=\pmatrix{U&Z\cr W&V\cr}\pmatrix{
P^I\cr i Q_I\cr},
\label{symptr}
\end{equation}
where the $4\times 4$ sub-matrices $U,V,W,Z$ have to satisfy the symplectic
constraints
\be
U^T V - W^T Z = V^T U - Z^T W = 1,\;\;\;
U^T W = W^T U,\;\;\;Z^T V = V^T Z.
\eq
Invariance of the lagrangian implies that $W=Z=0$, $VU^T=1$. In case
that $Z=0$, $W\neq 0$ and hence $VU^T=1$ the action is invariant up
to shifts in the $\theta$-angles; this is just the situation  one
encounters at the one-loop level. The non-vanishing matrix $W$
corresponds to non-trivial one-loop monodromy due to logarithmic
singularities in the prepotential. (This will be the subject of section 3.)
Finally, if $Z\neq 0$ then the electric fields transform into magnetic fields;
these transformations are the non-perturbative monodromies due to
logarithmic singularities induced by monopoles, dyons or other
non-perturbative excitations (see section 4).

The classical action is completely invariant under the target space
duality transformations. Thus the classical monodromies have $W,Z=0$.
The matrices $U$ (and hence $V=U^{T,-1}=U^*$) are given by
\be
U_{\sigma} = \left( \begin{array}{cccc}
1&0&0&0\\
0&1&0&0\\
0&0&0&1\\
0&0&1&0\\
\end{array} \right),\;\;\;
U_{g_1} = \left( \begin{array}{cccc}
0&0&1&0\\
0&0&0&1\\
-1&0&0&0\\
0&-1&0&0\\
\end{array} \right),\;\;\;
U_{g_2} = \left( \begin{array}{cccc}
1&0&1&0\\
0&0&0&1\\
-1&0&0&0\\
0&-1&0&1\\
\end{array} \right)
\eq
At the classical level the $S$-field is invariant under these
transformations.
The corresponding symplectic matrices for the four inequivalent
Weyl reflections  then immediately follow as
\beqa
U_{w_1} &=&\left( \begin{array}{cccc}
1&0&0&0 \\ 0 &1&0&0
\\ 0&0&0&1 \\  0&0&1&0 \\
\end{array} \right)   \;\;\;,\;\;\;
U_{w_2}=U_{w_1'}= \left( \begin{array}{cccc}
0&1&0 & 0 \\ 1&0&0&0 \\
0& 0&1&0 \\ 0&0&0&1 \\
\end{array} \right)   \nonumber\\
U_{w_2'} &=& \left( \begin{array}{cccc}
1&0&0&0 \\ 1&1&1&-1 \\
-1&0&0&1 \\ 1&0&1&0 \\
\end{array} \right)   \;\;\;,\;\;\;
U_{w_0'}=\left( \begin{array}{cccc}
1&1&1&-1 \\ 0&1&0&0 \\
0&-1&0&1 \\ 0&1&1&0 \\
\end{array} \right)
\label{umatrices}
\eeqa

Let us now discuss the masses of those states
which saturate the socalled BPS bound. These masses are dermined
by the complex central charge of the $N=2$ supersymmetry algebra.
In general the mass formula is given by the following expression
\cite{FKLZ,CAFP}
\begin{equation}
M^2=e^K|M_IP^I+ i N^IQ_I|^2=e^K|{\cal M}|^2.
\label{massn2}
\end{equation}
Here $K$ is the K\"ahler potential,
the $M_I$ are the electric quantum numbers of $U(1)^4$ and
the $N^I$ are the magnetic quantum numbers. It follows that the
classical spectrum of electric states, i.e. $N^I=0$, agrees with
the string momentum and winding spectrum of eq.(\ref{HMF}),
upon identification
$M_I=(m_2,-n_2,n_1,-m_1)$.
Moreover, if one chooses linearly dependent
electric and magnetic charges,
i.e. $M_I/p=(m_2,-n_2,n_1,-m_1), N^I/q=(-n_2,m_2,-m_1,n_1)$,
then the classical mass formula factorizes as\footnote{
We will, however, in the following not rely on this factorisation,
but rather use equation (\ref{massn2}).}
\begin{equation}
M^2={|(p+iqS)(m_2-im_1U+in_1T-n_2UT)|^2\over (S+\bar S)(T+\bar T)(U+\bar U)}.
\end{equation}
The moduli dependent part again agrees with the classical string momentum and
winding spectrum; $p$ and $q$ are the electric and magnetic
$S$-quantum numbers.
Note that the factorized classical mass formula can be obtained by
truncating the BPS mass formula of the $N=4$ heterotic string to
the $S,T,U$ subspace.

Finally, requiring the holomorphic mass ${\cal M}$ in (\ref{massn2})
to be invariant under the symplectic transformations (\ref{symptr})
yields that the quantum numbers $M_I$ and $N^I$ have to transform
as
\beqa
(N,-M)^T \rightarrow (N,-M)^T \; \Gamma^T
\eeqa
under (\ref{symptr}).

\section{Perturbative results \label{pertsec}}

\setcounter{equation}{0}

Let us first review the main results about the one-loop perturbative
holomorphic prepotential which were derived in \cite{WKLL,AFGNT}.
Using simple power counting arguments it is clear that the one-loop
prepotential must be independent of the dilaton field $S$. The same
kind of arguments actually imply that there are no higher loop corrections
to the prepotential in perturbation theory. Thus the perturbative, i.e.
one loop prepotential takes the form
\be
F = F^{(Tree)}(X) + F^{(1-loop)}(X)
= i \frac{X^1 X^2 X^3}{X^0} + (X^0)^2 f(T,U) = - STU + f(T,U)
\eq
where
\be
S = i \frac{X^1}{X^0},\;\;\;
T = -i \frac{X^2}{X^0},\;\;\;
U = -i \frac{X^3}{X^0}
\label{speccoord}
\eq
Taking derivatives of the 1-loop corrected prepotential gives that
\beqa
F_0 &=& -i \frac{X^1 X^2 X^3}{(X^0)^2} + 2 X^0 f(T,U) + i X^2 f_T +
i X^3 f_U \;,\;
F_1 = i\frac{X^2 X^3}{X^0},\nonumber\\
F_2 &=& i \frac{X^1 X^3}{X^0} -i X^0 f_T,\quad
F_3 = i \frac{X^1 X^2}{X^0} -i X^0 f_U
\eeqa
which, in special coordinates, turns into
\be
F_0 = STU + 2f(T,U) - T f_T - U f_U,\;\;\;
F_1 = -i TU,\;\;\;
F_2 = iSU - i f_T,\;\;\;
F_3 = iST - i f_U
\eq
The 1-loop corrected
K\"ahler potential is, in special coordinates, given by
\beqa
K(S,T,U) = - \log Y,\;\;\;Y = Y_{tree} + Y_{pert},\;\;\;
Y_{tree} = (S+\overline{S}) (T+\overline{T})(U+\overline{U})
\nonumber\\
Y_{pert} = 2(f + \overline{f}) - (T+\overline{T})
(f_T + \overline{f_T}) - (U+\overline{U})
(f_U + \overline{f_U})
\eeqa
The 1-loop corrected
section $\Omega^T = ( P, iQ)^T=(P^0,P^1,P^2,P^3,iQ_0,iQ_1,iQ_2,iQ_3)$
is given by
\beqa
\Omega^T &=& (X^0, iF_1, X^2, X^3, iF_0, -X^1, iF_2, iF_3) \nonumber\\
&=&
(1, TU, iT, iU, iSTU + 2i f - iT f_T - iU f_U, iS, -SU + f_T, -ST +
f_U)
\label{omegapert}
\eeqa

Since the target space duality transformations are known to be
a symmetry in each order of perturbation theory, the tree level plus
one-loop effective action must be invariant under these transformations, where
however one has to allow for discrete shifts in the various $\theta$
angles
due to monodromies around semi-classical
singularities in the moduli space where massive string modes become
massless.
Instead of the classical transformation rules,
in the quantum theory, $(P^I,i Q_I)$ transform according to
\begin{equation}
P^I\ \rightarrow\    U{}^I_{\,J}\, P^J,\qquad
i Q_I\ \rightarrow\    V_I{}^J \,i Q_J\ +\   W_{IJ}\, P^J\ ,
\label{symptrans}
\end{equation}
where
\begin{equation}
 V = ( U^{\rm T})^{-1},\quad
 W =  V \Lambda\,,\quad \Lambda=\Lambda^{\rm T}
\end{equation}
and $ U$ belongs to  $SO(2,2,{\bf Z})$.
Classically, $\Lambda=0$, but in the quantum theory, $\Lambda$
is a real symmetric matrix, which should be integer
valued in some basis.

Besides the target space duality symmetries, the effective action is
also invariant, up to discrete shifts in the $\theta$-angles,
under discrete shifts in the $S$-field, $D$: $S\rightarrow S+i$.
Thus the full perturbative monodromies contain the following $Sp(8,{\bf Z})$
transformation:
\be
V_{S} = U_S = \left( \begin{array}{cccc}
1&0&0&0\\
0&1&0&0\\
0&0&1&0\\
0&0&0&1\\
\end{array} \right),\;\;\;
W_S = \left( \begin{array}{cccc}
0&-1&0&0\\
-1&0&0&0\\
0&0&0&-1\\
0&0&-1&0\\
\end{array} \right),\;\;\; Z_S = 0.
\label{sdualpert}
\eq

Invariance of the one-loop action up to discrete $\theta$-shifts then implies
that
\be
F^{(1-loop)} (X) \longrightarrow F^{(1-loop)} (X) - \frac{i}{2}
\Lambda_{IJ} P^I P^J
\eq
This reads in special coordinates like
\be
f(T,U) \rightarrow (icT + d)^{-2} (f(T,U) + \Psi(T,U))
\eq
for an arbitrary $PSL(2,{\bf Z})_T$ transformation.
$\Psi(T,U)$ is a quadratic polynomial in $T$ and $U$.

As explained in \cite{WKLL,AFGNT}
the dilaton is not any longer invariant under
the target space duality transformations at the one-loop level.
Indeed, the relations (\ref{speccoord}) and
(\ref{symptrans}) imply\footnote{ It is still
possible to define an invariant dilaton field which is however
not an $N=2$ special coordinate \cite{WKLL}.}
\be
S \longrightarrow S + \frac{V_1^{\;J} (F^{(1-loop)}_J -i
\Lambda_{JK} P^K )}{U^{0}_{\;I} P^I}
\label{spert}
\eq

Near the singular lines the one-loop prepotential exhibits
logarithmic singularities and is therefore not a singlevalued function
when transporting the moduli fields around the singular lines.
For example around the singular $SU(2)_{(1)}$ line $T=U \neq 1,\rho$
the function $f$
must have the
following form
\begin{equation}
f(T,U)={1\over \pi}(T-U)^2\log(T-U)+\Delta(T,U),
\end{equation}
where $\Delta(T,U)$ is finite and single valued at $T=U \neq 1,\rho$.
At the remaining three critical  lines $f(T,U)$ takes an analogous form.
Moreover at the intersection points the residue of the
singularity must change in agreement with the number of states
which become massless at these critical points  (These residues are
of course just given by the $N=2$ pure Yang-Mills $\beta$-functions
for $SU(2)$, $SU(2)^2$ and $SU3)$ (there are no massless additional flavors
at the points of enhanced symmetries).)
Specifically at the point $(T,U)=(1,1)$ the
prepotential takes the form
\begin{equation}
f(T,U=1)={1\over \pi}(T-1)\log(T-1)^2+\Delta'(T)
\end{equation}
and around $(T,U)=(\rho,\bar\rho)$
\begin{equation}
f(T,U={\bar \rho})={1\over \pi}(T-\rho)\log(T-\rho)^3+\Delta''(T),
\end{equation}
where $\Delta'(T)$, $\Delta''(T)$ are finite at $T=1$, $T=\rho$ respectively.
Since $f(T,U)$ is not a true modular form, but has non-trivial monodromy
properties, it is not possible to determine the exact analytic
form of $f(T,U)$. However the third derivative transforms nicely
under target space duality transformtions, and using the informations
about the order of poles and zeroes one can uniquely determine
\[
\partial^3_T f(T,U) \propto  \frac{+1}{2 \pi}
\frac{E_4(iT) E_4(iU) E_6(iU) \eta^{-24}(iU)}{j(iT) - j(iU)}
\]
\be
\partial^3_U f(T,U) \propto \frac{-1}{2 \pi}
\frac{E_4(iT) E_6(iU) \eta^{-24}(iT) E_4(iU)}{j(iT) - j(iU)}
\label{ThirdDerivative}
\eq

This result has recently prooved to be important to support the
hypotheses \cite{KacVaf,many3}
that the quantum vector moduli space of the $N=2$ heterotic string
is given  by the tree level vector moduli space of an dual type II,
$N=2$ string, compactified on a suitably choosen Calabi-Yau space.
In addition to eq.(\ref{ThirdDerivative}) one can also deduce that
\cite{CLM,WKLL}
\be
\partial_T \partial_U f =- \frac{2}{\pi}
\log(j(iT) - j(iU)) + \mbox{finite}
\eq
which has precisely the right property that the coefficient
of the logarthmic singularity is proportional to the number
of generically massive states that become massless.

Using eqs. (\ref{massn2}) and (\ref{omegapert})
we can also determine the loop corrected mass formula
for the $N=2$ BPS states:
\beqa
{\cal M} &=&  M_0+M_1TU + iM_2T+iM_3U+iN^0(STU+2f-Tf_T-Uf_U) \nonumber\\
&+& i N^1 S+N^2(f_T-SU)+N^3(f_U-ST)
\eeqa
We recognize that electric states with $N^I=0$ do not get a mass shift
at the perturbative level. It follows that the positions of the singular
loci of enhanced gauge symmetries are unchanged in perturbation theory.
However the masses of states with magnetic charges
$N^I\neq 0$ are already shifted at the perturbative level.

\subsection{Perturbative $SU(2)_{(1)}$ monodromies}

Let us now consider the element $\sigma$
which corresponds to the Weyl reflection in the first enhanced $SU(2)_{(1)}$.

Under the
mirror transformation $\sigma$,
$T \leftrightarrow U, T-U \rightarrow e^{-i \pi}
(T-U)$, and the
$P$ transform classically and perturbatively as
\be
P^0 \rightarrow P^0,\;\;\;
P^1 \rightarrow P^1,\;\;\;
P^2 \rightarrow P^3,\;\;\;
P^3 \rightarrow P^2
\eq
The one--loop correction $f(T,U)$ transforms as\footnote{
Note that one can always add polynomials of quadratic order
in the moduli to a given $f(T,U)$ \cite{AFGNT}. This results
in the conjugation of the monodromy matrices.  Hence, all
the monodromy matrices given in the following are unique up to
conjugation.}
\beqa
f(T,U) &\rightarrow& f(U,T) = f(T,U) -i(T-U)^2 \nonumber\\
f_T(U,T) &=& f_U(T,U) + 2i (T-U),\;\;\;
f_U(U,T) = f_T(T,U) - 2i (T-U)
\eeqa
The $f$ function must then have the following form for
$T \rightarrow U$
\be
f(T,U) =  \frac{1}{\pi} (T-U)^2 \log(T-U) + \Delta(T,U)
\label{fisu2}
\eq
with derivatives
\beqa
f_T(T,U) =  \frac{2}{\pi} (T-U) \log(T-U) + \frac{1}{\pi} (T-U) +
\Delta_T
\nonumber\\
f_U(T,U) =  -\frac{2}{\pi} (T-U) \log(T-U) - \frac{1}{\pi} (T-U)
+ \Delta_U
\label{fsu2tu}
\eeqa
$\Delta(T,U)$ has the property that it is finite as $T \rightarrow U \neq
1,\rho$ and that, under mirror symmetry $T \leftrightarrow U$,
$\Delta_T \leftrightarrow \Delta_U$.
The 1-loop corrected $Q_2$ and $Q_3$ are thus given by
\beqa
Q_2 = i S U
- \frac{2i}{\pi} (T-U) \log(T-U) - \frac{i}{\pi} (T-U) -i \Delta_T \nonumber\\
Q_3 = i S T
+ \frac{2i}{\pi} (T-U) \log(T-U) + \frac{i}{\pi} (T-U)
-i \Delta_U
\label{q2q3}
\eeqa

It follows from (\ref{spert}) that,
under mirror symmetry $T \leftrightarrow U$,
the dilaton $S$ transforms as
\be
S \rightarrow S + i
\label{stransf}
\eq
Then, it follows that
perturbatively
\beqa
\left( \begin{array}{c}
Q_2\\  Q_3\\
\end{array} \right)
\rightarrow
\left( \begin{array}{c}
Q_3\\  Q_2\\
\end{array} \right) +
\left( \begin{array}{cc}
1 & -2\\
-2 & 1 \\
\end{array} \right)
\left( \begin{array}{c}
T\\  U\\
\end{array} \right)
\label{su2mono}
\eeqa
Thus,
the section $\Omega$ transforms perturbatively as
$\Omega \rightarrow \Gamma_{\infty}^{w_1} \Omega$, where
\beqa
\Gamma_{\infty}^{w_1} &=&
\left( \begin{array}{cc}
U & 0 \\ U \Lambda & U\\
\end{array} \right)  \;\;,\;\;
U=\left( \begin{array}{cc}
I & 0 \\ 0 &  \eta \\
\end{array} \right)   \;\;,\;\;
\Lambda =-
\left( \begin{array}{cc}
\eta & 0 \\
0 & {\cal C}\\
\end{array} \right) \nonumber\\
\eta &=& \left( \begin{array}{cc}
0& 1 \\
1 & 0\\
\end{array} \right) \;\;,\;\;
{\cal C} = \left( \begin{array}{cc}
2& -1 \\
-1 & 2\\
\end{array} \right)
\nonumber\\
\label{perttrans}
\eeqa

\subsection{Truncation to the rigid case of Seiberg/Witten \label{secseiwit}}

In order to truncate the perturbative $SU(2)_{(1)}$ monodromy
$\Gamma_{\infty}^{w_1}$
to the rigid one of Seiberg/Witten \cite{SW1}, we will take the limit
$\kappa^2=\frac{8 \pi}{M^2_{pl}} \rightarrow 0$ as well as expand
\beqa
T = T_0 + \kappa \delta T \nonumber\\
U= T_0 + \kappa \delta U
\eeqa
Here we have expanded the moduli fields $T$ and $U$ around the same
vev $T_0 \neq 1,\rho$.  Both $ \delta T$ and $\delta U$ denote
fluctuating fields of mass dimension one.  We will also freeze in the
dilaton field to a
large vev, that is we will set $S = \langle S \rangle \rightarrow
\infty$.
Then, the $Q_2$ and $Q_3$ given in (\ref{q2q3}) can be expanded
as
\beqa
Q_2 &=&  i \langle S \rangle T_0 + \kappa \tilde{Q}_2 \;\;\;,\;\;\;
Q_3 = i \langle S \rangle T_0 + \kappa \tilde{Q}_3 \nonumber\\
\tilde{Q}_2 &=& i \langle S \rangle \delta U
- \frac{2i}{\pi} (\delta T- \delta U)
\log\kappa^2 ( \delta T - \delta U) - \frac{i}{\pi} (\delta T- \delta U)
-i \Delta_T(\delta T,\delta U) \nonumber\\
\tilde{Q}_3 &=& i \langle S \rangle \delta T
+ \frac{2i}{\pi} (\delta T- \delta U)
\log\kappa^2 ( \delta T - \delta U) + \frac{i}{\pi} (\delta T- \delta U)
-i \Delta_U(\delta T,\delta U) \nonumber\\
\label{truncq2q3}
\eeqa
Next, one has to specify how mirror symmetry is to act on
the vev's $T_0$ and $\langle S \rangle$ as well as on $\delta T$ and
$\delta U$.  We will take that under mirror symmetry
\beqa
T_0 \rightarrow T_0 \;\;,\;\; \delta T \leftrightarrow \delta U \;\;,\;\;
\langle S \rangle \rightarrow
\langle S \rangle
\label{tSTU}
\eeqa
Note that we have taken $\langle S \rangle$ to be invariant under mirror
symmetry.  This is an important difference to (\ref{stransf}).
Using (\ref{tSTU}) and that $\delta T - \delta U \rightarrow
e^{-i \pi} (\delta T - \delta U)$, it follows that the truncated
quantities $\tilde{Q}_2$ and $\tilde{Q}_3$ transform as
follows under mirror symmetry
\beqa
\left( \begin{array}{c}
\tilde{Q}_2\\  \tilde{Q}_3\\
\end{array} \right)
\rightarrow
\left( \begin{array}{c}
\tilde{Q}_3\\  \tilde{Q}_2\\
\end{array} \right) +
\left( \begin{array}{cc}
2 & -2\\
-2 & 2 \\
\end{array} \right)
\left( \begin{array}{c}
\delta T\\ \delta  U\\
\end{array} \right)
\label{truncsu2mono}
\eeqa
Defing a truncated section $\tilde{\Omega}^T= (\tilde{P}^2,\tilde{P}^3,
i\tilde{Q}_2,i\tilde{Q}_3)=(i \delta T,i \delta U,
i\tilde{Q}_2,i\tilde{Q}_3)$, it follows that $\tilde{\Omega}$ transforms
as $\tilde{\Omega} \rightarrow \tilde{\Gamma}_{\infty}^{w_1} \tilde{\Omega}$
under mirror symmetry (\ref{tSTU}) where
\beqa
\tilde{\Gamma}_{\infty}^{w_1} =
\left( \begin{array}{cc}
\tilde{U} & 0 \\ \tilde{U} \tilde{\Lambda} & \tilde{U}\\
\end{array} \right)  \;\;,\;\;
\tilde{U}= \eta \;\;,\;\;
\eta = \left( \begin{array}{cc}
0& 1 \\
1 & 0\\
\end{array} \right) \;\;,\;\;
\tilde{\Lambda} = \left( \begin{array}{cc}
-2& 2 \\
2 & -2\\
\end{array} \right)
\label{truncmonotrans}
\eeqa
Note that, because of the invariance of $\langle S \rangle$ under mirror
symmetry, $\tilde{\Lambda} \neq - {\cal C}$, contrary to what one
would have gotten by performing a naive truncation of (\ref{perttrans})
consisting in keeping only rows and columns associated with
$(P^2,P^3,iQ_2,iQ_3)$.

Finally, in order to compare the truncated $SU(2)$ monodromy
(\ref{truncmonotrans}) with the perturbative $SU(2)$ monodromy
of Seiberg/Witten \cite{SW1}, one has to perform a change of basis from
moduli fields to Higgs fields, as follows
\beqa
\left( \begin{array}{c}
a \\
a_D\\
\end{array} \right)
=
M \tilde{\Omega} \;\;,\;\;
M =
\left( \begin{array}{cc}
m &  \\
 & m^* \\
\end{array} \right)  \;\;,\;\;
m= \frac{\gamma}{\sqrt{2}}
\left( \begin{array}{cc}
1& -1 \\
1& 1\\
\end{array} \right)
\label{higgsbasis}
\eeqa
where $\gamma$ denotes a constant to be fixed below.  Then, the
perturbative $SU(2)$ monodromy in the Higgs basis is given by
\beqa
\tilde{\Gamma}^{Higgs}_{\infty} = M \tilde{\Gamma}_{\infty}^{w_1} M^{-1}
=
\left( \begin{array}{cc}
m \tilde{U} m^{-1}  &   0  \\
m^* \tilde{U} \tilde{\Lambda} m^{-1} &
m^* \tilde{U} m^T \\
\end{array} \right)
\eeqa
which is computed to be
\beqa
\tilde{\Gamma}^{Higgs,w_1}_{\infty} =
\left( \begin{array}{cccc}
-1 & & & \\
& 1 & & \\
\frac{4}{\gamma^2} & 0 & -1 & 0 \\
& & & 1\\
\end{array} \right)
\label{higgsmono}
\eeqa
Note that (\ref{higgsmono}) indeed correctly shows that, under the
Weyl reflection in the first $SU(2)$, the second $SU(2)$ is left untouched.
The fact that (\ref{higgsmono}) reproduces this behaviour
can be easily traced back to the fact that we have assumed that $
\langle S \rangle$ stays invariant under the mirror transformation
$\delta T \leftrightarrow \delta U$.  Finally, comparing with the
perturbative $SU(2)$ monodromy of Seiberg/Witten \cite{SW1} yields that
$\gamma^2 = 2$, whereas comparision with the perturbative $SU(2)$
monodromy of Klemm et al \cite{KLT} gives that $\gamma^2 = 1$.

\subsection{Relating $\Lambda$ to the dilaton vev $\langle S \rangle $
\label{lamdil}}

In the following we will consider the rigid limit and
relate the dynamically generated scale $\Lambda$
of Seiberg/Witten \cite{SW1} to the frozen dilaton vev $\langle S \rangle $.

We took the
$f$ function to be of the following form for
$T \rightarrow U$
\be
f(T,U) =  \frac{1}{\pi} (T-U)^2 \log(T-U) + \Delta(T,U)
\eq
$\Delta (T,U)$ denotes a 1-loop contribution coming from additional
heavy modes associated with $SU(2)_{(2)}$.  For energies $E^2$ in a regime
where $|\delta T + \delta U|^2 \gg
|\delta T-\delta U|^2 \gg E^2 \gg \Lambda^2$,
these heavier modes decouple from the
low energy effective action and
the 1-loop correction  is due to the light modes associated with the
first $SU(2)_{(1)}$, only.  Then, in this regime the 1-loop contribution
$\Delta(T,U)$ can be safely ignored.

The Higgs section $(a,a_D)^T=(a_1,a_2,a_{D1},a_{D2})$
is obtained from the truncated section
$\tilde{\Omega}^T= (\tilde{P}^2,\tilde{P}^3,
i\tilde{Q}_2,i\tilde{Q}_3)=(i \delta T,i \delta U,
i\tilde{Q}_2,i\tilde{Q}_3)$
via
\beqa
\left( \begin{array}{c}
a \\
a_D\\
\end{array} \right)
=
 M \tilde{\Omega} \;\;,\;\;
M =
\left( \begin{array}{cc}
m &  \\
 & m^* \\
\end{array} \right)  \;\;,\;\;
m= \frac{\gamma}{\sqrt{2}}
\left( \begin{array}{cc}
1& -1 \\
1& 1\\
\end{array} \right)
\eeqa
Then
\beqa
a_1 &=& \frac{i\gamma}{\sqrt{2}} \left( \delta T - \delta U \right) \;\;,\;\;
a_2 = \frac{i\gamma}{\sqrt{2}} \left( \delta T + \delta U \right) \nonumber\\
a_{D1} &=& \frac{i}{\sqrt{2}\gamma} \left( \tilde{Q}_2 - \tilde{Q}_3 \right)
\nonumber\\
&=&  \frac{1}{\sqrt{2}\gamma} \left[ \langle S \rangle
\left( \delta T - \delta U \right)
+ \frac{4}{\pi} \left( \delta T - \delta U \right) \log \left(
\delta T - \delta U
\right)
+   \frac{2}{\pi} \left( \delta T - \delta U \right) \right]
\nonumber\\
a_{D2} &=&  \frac{i}{\sqrt{2}\gamma} \left( \tilde{Q}_2 +\tilde{Q}_3 \right)
= - \frac{1}{\sqrt{2}\gamma} \langle  S \rangle
\left( \delta T + \delta U \right)
\eeqa
and consequently
\beqa
a_{D1} &=& - \frac{i}{\gamma^2}
\langle S \rangle a_1  - \frac{4i}{\pi\gamma^2} a_1 \log
\left( \frac{\sqrt{2}}{\gamma} a_1\right) - \frac{2 i }{\pi \gamma^2}
a_1 - \frac{2}{\gamma^2} a_1 \nonumber\\
&=& \frac{i}{\gamma^2} a_1 \left( - \langle S \rangle   - \frac{4}{\pi}  \log
\left( \frac{\sqrt{2}}{\gamma} a_1\right) - \frac{2 }{\pi}  + 2i \right)
\nonumber\\
a_{D2} &=& \frac{i}{\gamma^2} \langle S \rangle  a_2
\eeqa
Setting\footnote{Seiberg/Witten corresponds to $\gamma^2 = 2$.
Taking
into account that their looping around singular points is opposite
to ours gives total agreement between our and their results.}
\beqa
a_{D1}=
-\frac{4i}{\pi \gamma^2}  a_1\log
\left( \frac{a_1}{\Lambda} \right) - \frac{2i}{\pi \gamma^2} a_1
\eeqa
it follows that
\beqa
\Lambda = e^{- \frac{\pi}{4} \langle S \rangle
- \log \frac{\sqrt{2}}{\gamma} + \frac{i \pi}{2}}
\eeqa
in the rigid case.

In the local case, on the other hand, the dynamically generated scale
\beqa
\Lambda = e^{- \frac{\pi}{4}  S
- \log \frac{\sqrt{2}}{\gamma} + \frac{i \pi}{2}}
\eeqa
is in general not invariant under modular transformations due
to an associated transformation of the dilaton $S$.

\subsection{Perturbative $SU(2)_{(2)}$ monodromies}

Under the Weyl twist $w_2$ in the
second $SU(2)_{(2)}$, the moduli $T$ and $U$ transform
as $T \rightarrow \frac{1}{U}, U \rightarrow \frac{1}{T}$.
The section $\Omega$ transforms perturbatively as
$\Omega \rightarrow \Gamma_{\infty}^{w_2} \Omega$.  $\Gamma_{\infty}^{w_2}$
is conjugated to $\Gamma_{\infty}^{w_1}$ by $\Gamma^{(g_1)}$.
Since $\Gamma^{(g_1)}$ can be taken to have no
perturbative corrections \cite{AFGNT}, we get that
\beqa
\Gamma_{\infty}^{w_2} &=&
\left( \begin{array}{cc}
U & 0 \\ U \Lambda & U\\
\end{array} \right)  \;,\;
U= \left( \begin{array}{cc}
\eta & 0 \\ 0 &  I \\
\end{array} \right)   \;,\;
\Lambda =
\left( \begin{array}{cc}
- {\cal C} & 0 \\
0 & -\eta\\
\end{array} \right) \nonumber\\
\eta &=& \left( \begin{array}{cc}
0& 1 \\
1 & 0\\
\end{array} \right) \;\;,\;\;
{\cal C} = \left( \begin{array}{cc}
2& -1 \\
-1 & 2\\
\end{array} \right)
\eeqa
It then follows that
perturbatively
\beqa
\left( \begin{array}{c}
Q_2\\  Q_3\\
\end{array} \right)
\rightarrow
\left( \begin{array}{c}
 Q_2\\   Q_3\\
\end{array} \right) -
\left( \begin{array}{c}
U\\  T\\
\end{array} \right)
\label{secsu2mono}
\eeqa

Next, let us construct 1-loop corrected $Q_2$ and $Q_3$ which have the
above monodromy properties.
We will show that the
1-loop correction $f(T,U)$ reproducing the
perturbative monodromy (\ref{secsu2mono}) is, in the vicinity of
$T=\frac{1}{U}$, given by
\beqa
f(T,U) &=& - \frac{1}{\pi}
(\delta T + \delta U)^2 \log(\delta T + \delta U) + \Xi (T,U)
\label{secsu2}
\eeqa
where
we have expanded $T=T_0( 1 + \delta T), U =\frac{1}{T_0}( 1 + \delta U)$.
$\Xi (T,U)$ and its derivatives $\Xi_{T,U}$ have the
property that $\Xi \rightarrow \Xi,
\Xi_{T,U} \rightarrow \Xi_{T,U}$
under the linearised transformation laws
$ \delta T \rightarrow - \delta U, \delta U
\rightarrow - \delta T,
\delta T + \delta U \rightarrow
e^{-i\pi}(\delta T  +\delta U)$.  An example of a $\Xi(T,U)$ meeting
these requirements is given by
$\Xi =  \frac{1}{\pi} (\delta T- \delta U)^2 \log(\delta T-\delta U)$.
Using (\ref{secsu2}), it follows that $Q_2$ and $Q_3$ are at the linearised
level given by
\beqa
Q_2 &=& i S U P^0 - i P^0 f_T \nonumber\\
&=&
i S \frac{(1 + \delta U)}{T_0}
- i \left(
- \frac{2}{\pi} \frac{(\delta T +  \delta U)}{T_0} \log(\delta T +  \delta U)
  -\frac{1}{\pi} \frac{(\delta T + \delta U)}{T_0} + \Xi_T
\right)
 \nonumber\\
Q_3 &=& i S T P^0 - i P^0 f_U \nonumber\\
&=&
i S T_0 (1 + \delta T)   -i
\left( -
\frac{2}{\pi} T_0 (\delta T+\delta U) \log(\delta T + \delta U)
- \frac{1}{\pi} T_0 ( \delta T + \delta U) + \Xi_U \right) \nonumber\\
\label{q2q3secsu2}
\eeqa
Now, under $T \rightarrow \frac{1}{U}, U \rightarrow \frac{1}{T}$,
the dilaton transforms as $S \rightarrow S -i + \frac{2i}{TU}
+ \frac{1}{TU} ( 2 f - T f_T - U f_U)$, whereas
the graviphoton transforms as $P^0 \rightarrow P^1$.
Linearising these transformation laws,
using the properties of $\Xi_{T,U}$ given above as well as
\beqa
2 f - T f_T - U f_U =  \frac{2}{\pi}
\left( \delta T + \delta U + 2 (\delta T + \delta U) \log (\delta T
+ \delta U) \right)
\label{fidentity}
\eeqa
gives that
\beqa
Q_2 &\rightarrow&  i S \frac{(1 + \delta U)}{T_0}
- i \left( -
\frac{2}{\pi}
  \frac{(\delta T + \delta U)}{T_0} \log (\delta T + \delta U)
- \frac{1}{\pi} \frac{(\delta T + \delta U)}{T_0} + \Xi_T \right)
- \frac{(1+\delta U)}{T_0}
 \nonumber\\
&=&  Q_2 - \frac{(1+\delta U)}{T_0}
\eeqa
and similarly that
\beqa
Q_3 \rightarrow  Q_3 - T_0 (1+\delta T)
\eeqa
Thus, the 1-loop correction $f$ given in (\ref{secsu2}) correctly
reproduces the perturbative monodromy (\ref{secsu2mono}).
Note that (\ref{fidentity}) implies that $2 \Xi -T \Xi_T -U \Xi_U =0$
which is an independent constraint on $\Xi$. Again this requirement
is satisfied by
\be
\Xi =  \frac{1}{\pi} (\delta T- \delta U)^2 \log(\delta T-\delta U)
\label{Xiconstraint}
\eq
at the linearized level.

\subsection{Truncation to the rigid case}

Next, consider truncating
(\ref{secsu2mono}) to the rigid case. In the rigid case
one expects to recover a second copy of the $SU(2)$-case discussed by
Seiberg/Witten \cite{SW1}.  In order to do so, we will freeze in both the
graviphoton $\langle P^0
\rangle =1$ and the dilaton $\langle S
\rangle =\infty$.  That is, both $P^0$
and $S$ will be taking to be invariant under $\delta T\rightarrow -
\delta U, \delta U \rightarrow - \delta T$.
Note that, in particular, $\langle P^0 \rangle= 1$ is a fixed point of $P^0
\rightarrow P^1 =  TU = (1 + \delta T + \delta U)$ in the local
case.

Then, (\ref{q2q3secsu2}) can be written as
\beqa
Q_2 &=& i \langle S \rangle \frac{1}{T_0} + \frac{1}{T_0} \tilde{Q}_2
\;\;,\;\;
Q_3 = i \langle S \rangle T_0 + T_0 \tilde{Q}_3 \nonumber\\
\tilde{Q}_2 &=& i \langle S \rangle \delta U
-i \left( -\frac{2}{\pi} (\delta T +  \delta U) \log(\delta T +  \delta U)
  - \frac{1}{\pi} ( \delta T + \delta U) + \Xi_{\delta T}
\right)
\nonumber\\
\tilde{Q}_3 &=& i \langle S \rangle \delta T
-i \left( -\frac{2}{\pi}(\delta T+\delta U) \log(\delta T + \delta U)
- \frac{1}{\pi} ( \delta T + \delta U) + \Xi_{\delta U}
\right)
\eeqa
Let as
impose yet another condition on $\Xi_{T,U}$, namely that
$\Xi_{\delta T} = - \Xi_{\delta U}$ at the linearised level.  Note that
$\Xi =  \frac{1}{\pi} (\delta T- \delta U)^2
\log(\delta T-\delta U)$ is an example of a $\Xi$ meeting
this additional requirement.
Then, it follows that under $\delta T \rightarrow - \delta U,
\delta T \rightarrow - \delta U , \delta T + \delta U
\rightarrow e^{-i\pi} ( \delta T + \delta U)$
\beqa
\left( \begin{array}{c}
\tilde{Q}_2\\  \tilde{Q}_3\\
\end{array} \right)
\rightarrow
\left( \begin{array}{c}
- \tilde{Q}_3\\  - \tilde{Q}_2\\
\end{array} \right) -  2
\left( \begin{array}{c}
\delta T + \delta U\\  \delta T + \delta U\\
\end{array} \right)
\eeqa
Thus, the truncated section ${\tilde \Omega}^T
= ( i \delta T, i \delta U, i \tilde{Q}_2,
i \tilde{Q}_3)$ transforms as
$\tilde{\Omega} \rightarrow \tilde{\Gamma}_{\infty}^{w_2} \tilde{\Omega}$
where
\beqa
\tilde{\Gamma}_{\infty}^{w_2} =
\left( \begin{array}{cc}
-\eta & 0 \\  -2I - 2 \eta & -\eta\\
\end{array} \right)  \;\;,\;\;
\eta = \left( \begin{array}{cc}
0& 1 \\
1 & 0\\
\end{array} \right)
\eeqa

The two critical lines $T=U$ and $TU=1$ are in the local
case related by the group element $g_1$ which acts
by $T \rightarrow 1/T$, $U \rightarrow U$. The monodromy
matrices associated with the two lines are related through
conjugation by $\Gamma^{(g_1)}$. This transformation permutes
the two $SU(2)$ factors (outer automorphism of $SU(2)^2$)
and therefore is also present in the rigid theory. Consequently
we expect that the two truncated monodromies
$\wt{\Gamma}_{\infty}^{(w_i)}$ are also conjugated.
The conjugation matrix is then the  truncated
version of $\Gamma^{(g_1)}$. Now the linearized
section transforms classically under $g_1$ by
\be
(i \delta T, i \delta U, i \wt{Q}_2, i \wt{Q}_3)
\rightarrow
(-i \delta T, i \delta U, i \wt{Q}_2, -i \wt{Q}_3)
\eq
Since $\Gamma^{(g_1)}$
has been taken to have no perturbative corrections \cite{AFGNT}, we
expect that this transformation law is likewise not
modified in the rigid case. Then
\be
\wt{\Gamma}^{(g_1)} = \left( \begin{array}{cccc}
-1&0&0&0\\
0&1&0&0\\
0&0&1&0\\
0&0&0&-1\\
\end{array} \right)
= (\wt{\Gamma}^{(g_1)})^{-1}
\eq
should be the truncated version of the permutation of the
two $SU(2)$s. And indeed one easily verifies that
$\wt{\Gamma}_{\infty}^{(w_2)} = \wt{\Gamma}^{(g_1)}
\wt{\Gamma}_{\infty}^{(w_1)} \wt{\Gamma}^{(g_1)}$.

\subsection{Perturbative $SU(2)^2$ monodromies}

Under the combined Weyl twists $w_1 w_2$
of $SU(2)^2$, the moduli $T$ and $U$ transform
as $T \rightarrow \frac{1}{T}, U \rightarrow \frac{1}{U}$.
The section $\Omega$ transforms perturbatively as
$\Omega \rightarrow \Gamma_{\infty}^{SU(2)^2} \Omega$, where
\beqa
\Gamma_{\infty}^{SU(2)^2} = \Gamma_{\infty}^{w_1} \Gamma_{\infty}^{w_2}=
\left( \begin{array}{cc}
U & 0 \\ X & U\\
\end{array} \right)  ,
U= \left( \begin{array}{cc}
\eta & 0 \\ 0 &  \eta \\
\end{array} \right)   ,
X = -2
\left( \begin{array}{cc}
\eta & 0 \\
0 & \eta \\
\end{array} \right) ,
\eta = \left( \begin{array}{cc}
0& 1 \\
1 & 0\\
\end{array} \right)
\eeqa
It then follows that
perturbatively
\beqa
\left( \begin{array}{c}
Q_2\\  Q_3\\
\end{array} \right)
\rightarrow
\left( \begin{array}{c}
 Q_3\\   Q_2\\
\end{array} \right) -  2
\left( \begin{array}{c}
U\\  T\\
\end{array} \right)
\label{su22mono}
\eeqa

Inspection of (\ref{fisu2}) and of (\ref{secsu2}) shows
that the
1-loop correction $f(T,U)$ reproducing the
perturbative monodromy (\ref{su22mono}) should, in the vicinity of
$T=U=1$, be given by
\beqa
f(T,U) =
\frac{1}{\pi} \left(
(\delta T- \delta U)^2 \log(\delta T- \delta U) -
(\delta T + \delta U)^2 \log(\delta T + \delta U) \right)
\label{ftu1}
\eeqa
Note that (\ref{ftu1}) satisfies all the requirements imposed on
$\Delta (T,U)$ and on $\Xi(T,U)$ in the previous sections.
Using (\ref{ftu1}), it follows that $Q_2$ and $Q_3$ are at the linearised
level given by
\beqa
Q_2 &=& i S U P^0 - i P^0 f_T \nonumber\\
&=&
i S (1 + \delta U)
-
\frac{2i}{\pi} \left(
(\delta T- \delta U) \log(\delta T- \delta U)
-(\delta T +  \delta U) \log(\delta T +  \delta U)
  - \delta U
\right)
 \nonumber\\
Q_3 &=& i S T P^0 - i P^0 f_U \nonumber\\
&=&
i S (1 + \delta T)   -
\frac{2i}{\pi}\left( -(\delta T-\delta U) \log(\delta T- \delta U)
-(\delta T+\delta U) \log(\delta T + \delta U)
- \delta T \right) \nonumber\\
\label{q2q3su22}
\eeqa
Under $w_1 w_2$,  $T \rightarrow \frac{1}{T}, U \rightarrow \frac{1}{U}$,
and the dilaton transforms as $S \rightarrow S + \frac{2i}{TU}
+ \frac{1}{TU} ( 2 f - T f_T - U f_U)$, whereas
the graviphoton transforms as $P^0 \rightarrow P^1$.
Linearising these transformation laws und using that
\beqa
2 f - T f_T - U f_U =  \frac{2}{\pi}
\left( \delta T + \delta U + 2 (\delta T + \delta U) \log (\delta T
+ \delta U) \right)
\eeqa
it follows that under $\delta T \rightarrow - \delta T = e^{-i\pi} \delta T,
\delta U \rightarrow - \delta U=e^{-i\pi} \delta U$
\beqa
Q_2 &\rightarrow&  i S (1 + \delta T)
- \frac{2i}{\pi} \left( - (\delta T- \delta U) \log(\delta T- \delta U)
-  (\delta T + \delta U) \log (\delta T + \delta U)  - \delta T\right)
\nonumber\\
&-& 2(1+\delta U)
=  Q_3 - 2(1+\delta U)
\eeqa
and similarly that
\beqa
Q_3 \rightarrow  Q_2 - 2 (1+\delta T)
\eeqa
Thus, the 1-loop correction $f$ given in (\ref{ftu1}) indeed correctly
reproduces the perturbative monodromy (\ref{su22mono}).

\subsection{Truncation to the rigid case}

Next, consider truncating the above $SU(2)^2$ monodromies
to the rigid case. In the rigid case
one expects to recover 2 copies of the $SU(2)$-case discussed by
Seiberg/Witten.  As before, we will freeze in both the
graviphoton and the dilaton to its fixed point values, i.e.
$\langle P^0
\rangle =1$, $\langle S
\rangle =\infty$.

Then, (\ref{q2q3su22}) can be written as
\beqa
Q_2 &=& i \langle S \rangle + \tilde{Q}_2 \;\;\;\;,\;\;\;\;
Q_3 = i \langle S \rangle + \tilde{Q}_3 \nonumber\\
\tilde{Q}_2 &=& i \langle S \rangle \delta U -
\frac{2i}{\pi} \left(
(\delta T- \delta U) \log(\delta T- \delta U)
-(\delta T +  \delta U) \log(\delta T +  \delta U)
  - \delta U
\right)
\nonumber\\
\tilde{Q}_3 &=& i \langle S \rangle \delta T -
\frac{2i}{\pi}\left( -(\delta T-\delta U) \log(\delta T- \delta U)
-(\delta T+\delta U) \log(\delta T + \delta U)
- \delta T \right) \nonumber\\
\eeqa
Then, it follows that under $\delta T \rightarrow e^{-i\pi} \delta T = -
\delta T, \delta U \rightarrow e^{-i\pi} \delta U = -
\delta U$
\beqa
\left( \begin{array}{c}
\tilde{Q}_2\\  \tilde{Q}_3\\
\end{array} \right)
\rightarrow
\left( \begin{array}{c}
- \tilde{Q}_2\\  - \tilde{Q}_3\\
\end{array} \right) -  4
\left( \begin{array}{c}
\delta U\\  \delta T\\
\end{array} \right)
\eeqa
Consequently,
the truncated section ${\tilde \Omega}$ transforms as
$\tilde{\Omega} \rightarrow \tilde{\Gamma}_{\infty}^{SU(2)^2} \tilde{\Omega}$
with
\beqa
\tilde{\Gamma}_{\infty}^{SU(2)^2} =
\left( \begin{array}{cc}
- I& 0 \\  - 4 \eta & -I \\
\end{array} \right)  \;\;,\;\;
\eta = \left( \begin{array}{cc}
0& 1 \\
1 & 0\\
\end{array} \right)
\eeqa
It can be checked that
\beqa
\tilde{\Gamma}_{\infty}^{SU(2)^2} =
\tilde{\Gamma}_{\infty}^{w_1}
\tilde{\Gamma}_{\infty}^{w_2}
\eeqa
as it must for consistency.
Finally, rotating to the Higgs basis gives that
\beqa
\tilde{\Gamma}_{\infty}^{Higgs,SU(2)^2} = M
\tilde{\Gamma}_{\infty}^{SU(2)^2} M^{-1} =
\left( \begin{array}{cccc}
-1 & & & \\
& -1 & & \\
\frac{4}{\gamma^2} & 0 & -1 & 0 \\
&- \frac{4}{\gamma^2} & & -1\\
\end{array} \right)
\eeqa

\subsection{The first Weyl twist $w_1'$ of $SU(3)$}

Under the first Weyl twist $w_1'$
of $SU(3)$, the moduli $T$ and $U$ transform
as $T \rightarrow \frac{1}{U}, U \rightarrow \frac{1}{T}$.
The section $\Omega$ transforms perturbatively as
$\Omega \rightarrow \Gamma_{\infty}^{w_1'} \Omega$, where
\beqa
\Gamma_{\infty}^{w_1'} &=& \Gamma_{\infty}^{w_2} =
\left( \begin{array}{cc}
U & 0 \\ U \Lambda & U\\
\end{array} \right)  \;,\;
U= \left( \begin{array}{cc}
\eta & 0 \\ 0 &  I \\
\end{array} \right)   \;,\;
\Lambda =
\left( \begin{array}{cc}
- {\cal C} & 0 \\
0 & -\eta \\
\end{array} \right) \nonumber\\
\eta &=& \left( \begin{array}{cc}
0& 1 \\
1 & 0\\
\end{array} \right) \;,\;
{\cal C}=
\left( \begin{array}{cc}
2& -1 \\
-1 & 2\\
\end{array} \right)
\label{monosu3w1}
\eeqa
It then follows that
perturbatively
\beqa
\left( \begin{array}{c}
Q_2\\  Q_3\\
\end{array} \right)
\rightarrow
\left( \begin{array}{c}
Q_2\\   Q_3\\
\end{array} \right) -
\left( \begin{array}{c}
U\\  T\\
\end{array} \right)
\label{su3mono}
\eeqa

Next, let us construct 1-loop corrected $Q_2$ and $Q_3$ which have the
above monodromy properties.
We will show that the
1-loop correction $f(T,U)$ reproducing the
perturbative monodromy (\ref{su3mono}) is, in the vicinity of
$T=\rho = \frac{1}{2} \sqrt{3} + \frac{i}{2}, U=\rho^{-1}$
given by
\beqa
f(T,U) = - \frac{1}{2 \pi} \left( \sum_i Z_i^2 \log Z_i
- \frac{1}{2} \sum_i Z_i^2 \right)\;\;,\;\; i=1,2,3
\label{fsu3}
\eeqa
where
\beqa
Z_1 &=& c \left( ( 2 - \rho^2) \delta T + (2 - \rho^{-2}) \delta U \right)
\nonumber\\
Z_2 &=& c \left( ( 2 \rho^2 -1) \delta T + (2  \rho^{-2}-1) \delta U \right)
\nonumber\\
Z_3 &=& c \left( ( \rho^2 +1) \delta T + (\rho^{-2} + 1) \delta U \right)
\label{z1z23z3}
\eeqa
and
where
we have expanded $T= \rho + \delta T, U = \rho^{-1} + \delta U$.
$c$ denotes a constant which can be determined as follows.  Differentiation
of (\ref{fsu3}) gives that
\beqa
f_{TU} (\delta T,\delta U)
=- \frac{3 c^2}{\pi} \left( \log Z_1 + \log Z_2 + \log Z_3 \right)
+ finite
\eeqa
It follows that
\beqa
f_{TU} (\delta T,\delta U=0)
= - \frac{9 c^2}{\pi}  \log \delta T
+ finite
\label{ftusu3}
\eeqa
The logarithmic singularity (\ref{ftusu3}) should be 3 times as
strong as the logarithmic singularity of the $SU(2)_1$ case given
by $f_{TU} ( T = U +
\delta T,U ) =- \frac{2}{\pi} \log \delta T$ as computed from (\ref{fsu2tu}).
Thus it follows that $c^2 = \frac{2}{3}$.
Using (\ref{fsu3}), it follows that $Q_2$ and $Q_3$ are at the linearised
level given by
\beqa
Q_2 &=& i S U P^0 - i P^0 f_T \nonumber\\
&=&
i S (\rho^{-1} + \delta U) + \frac{ci}{\pi} \left(
(2- \rho^2) Z_1 \log Z_1 + (2 \rho^2 -1) Z_2 \log Z_2
+ (\rho^2 + 1) Z_3 \log Z_3  \right)
 \nonumber\\
Q_3 &=& i S T P^0 - i P^0 f_U \nonumber\\
&=&
i S (\rho + \delta T)   + \frac{ci}{\pi} \left(
(2 - \rho^{-2}) Z_1 \log Z_1 + (2 \rho^{-2} - 1) Z_2 \log Z_2
+ (\rho^{-2} + 1 ) Z_3 \log Z_3 \right)
\nonumber\\
\label{q2q3su3}
\eeqa
Now, under $T \rightarrow \frac{1}{U}, U \rightarrow \frac{1}{T}$,
the dilaton transforms as $S \rightarrow S -i + \frac{2i}{TU}
+ \frac{1}{TU} ( 2 f - T f_T - U f_U)$, whereas
the graviphoton transforms as $P^0 \rightarrow P^1$.
Also, at the linearised level, $\delta T \rightarrow - \rho^2
\delta U, \delta U \rightarrow - \rho^{-2} \delta T$ and, consequently,
$Z_1 \rightarrow  e^{-i \pi} Z_1 = - Z_1, Z_2 \leftrightarrow Z_3$.
Using that $c^2=\frac{2}{3}$ and that
\beqa
2 f - T f_T - U f_U =  \frac{c \sqrt{3}}{\pi} \left(
2 Z_1 \log Z_1 - Z_2 \log Z_2 + Z_3 \log Z_3 \right) + {\cal O} (\delta T
\delta U)
\eeqa
it follows that at the linearised level $Q_2$ transforms into
\beqa
Q_2 &\rightarrow&  i S (\rho^{-1} + \delta U)
- \frac{ci}{\pi} \left( (\frac{3}{2} - \frac{i}{2}\sqrt{3}) Z_1 \log Z_1
+ i \sqrt{3} Z_2 \log Z_2 + ( \frac{3}{2} + \frac{i}{2}\sqrt{3} )
Z_3 \log Z_3  \right) \nonumber\\
&-& ( \rho^{-1} + \delta U)
=  Q_2 - U
\eeqa
and similarly that
\beqa
Q_3 \rightarrow  Q_3 - T
\eeqa
Thus, the 1-loop correction $f$ given in (\ref{fsu3}) correctly
reproduces the perturbative monodromy (\ref{su3mono}).
Note that, although the 1-loop correction $f$ given in (\ref{fsu3})
differs radically from the $f$ function for the $SU(2)_2$ case
given in (\ref{secsu2}), both 1-loop $f$ functions nevertheless give
rise to the same perturbative monodromy matrix
(\ref{monosu3w1}).  This is a consequence of the nontrivial transformation
law of the dilaton.

\subsection{Truncation to the rigid case \label{su3rig}}

Next, consider truncating the above to the rigid case.
In order to do so, we will freeze in both the
graviphoton and the dilaton to its fixed point values, i.e.
$\langle P^0
\rangle =1$ and $\langle S
\rangle =\infty$.  In order to compare the truncated $SU(3)$ monodromies
with the rigid monodromies of Klemm et al \cite{KLT}, one has to perform
a change of basis from the moduli fields to the Higgs fields.
The Higgs section $(a,a_D)^T=(a_1,a_2,a_{D1},a_{D2})$ is obtained
from the truncated section ${\tilde \Omega}^T = (i \delta T,i \delta U,
i {\tilde Q}_2 , i {\tilde Q}_3)$
via
\beqa
\left( \begin{array}{c}
a \\  a_D\\
\end{array} \right)=
\left( \begin{array}{cc}
m &  \\
& m^* \\
\end{array} \right)
\left( \begin{array}{c}
{\tilde P}\\ i {\tilde Q} \\
\end{array} \right) \;\;,\;\;
m= -i c
\left( \begin{array}{cc}
1 & 1 \\
\rho^{2}& \rho^{-2} \\
\end{array} \right) \;\;,\;\; c^2=\frac{2}{3}
\label{su3higgs}
\eeqa
Then, indeed,
\beqa
Z_1 &=& c \left( ( 2 - \rho^2) \delta T + (2 - \rho^{-2}) \delta U \right)
=2 a_1 - a_2
\nonumber\\
Z_2 &=& c \left( ( 2 \rho^2 -1) \delta T + (2  \rho^{-2}-1) \delta U \right)
=2 a_2 - a_1
\nonumber\\
Z_3 &=& c \left( ( \rho^2 +1) \delta T + (\rho^{-2} + 1) \delta U \right)
= a_1 + a_2
\eeqa
thus precisely reproducing equation (3.9) of Klemm et al \cite{KLT}.

Equation (\ref{q2q3su3}) can now be written as
\beqa
Q_2 = i \langle S \rangle \rho^{-1} + \tilde{Q}_2 \;\;,\;\;
Q_3 = i \langle S \rangle \rho + \tilde{Q}_3
\eeqa
with
\beqa
\tilde{Q}_2 &=& \frac{1}{\sqrt{3}c} \langle S \rangle (-a_2 + \rho^{2} a_1)
\nonumber\\
&+&\frac{ci}{\pi} \left(
(2- \rho^2) Z_1 \log Z_1 + (2 \rho^2 -1) Z_2 \log Z_2
+ (\rho^2 + 1) Z_3 \log Z_3  \right)
 \nonumber\\
\tilde{Q}_3 &=& \frac{1}{\sqrt{3}c} \langle S \rangle ( a_2 - \rho^{-2} a_1)
\nonumber\\
&+& \frac{ci}{\pi} \left(
(2 - \rho^{-2}) Z_1 \log Z_1 + (2 \rho^{-2} - 1) Z_2 \log Z_2
+ (\rho^{-2} + 1 ) Z_3 \log Z_3 \right)
\eeqa
Then, using (\ref{su3higgs}) it follows that
\beqa
a_{D1} &=& \frac{i}{2} \langle S \rangle (a_2 -2 a_1)
-\frac{i}{\pi} \left(2 Z_1 \log Z_1 - Z_2 \log Z_2 + Z_3 \log Z_3 \right)
\nonumber\\
a_{D2} &=&  \frac{i}{2} \langle S \rangle (a_1-2a_2)
-\frac{i}{\pi} \left(- Z_1 \log Z_1 + 2 Z_2 \log Z_2 + Z_3 \log Z_3 \right)
\label{higgsad}
\eeqa
Writing
\beqa
a_{D1} &=&
-\frac{i}{\pi} \left(2 Z_1 \log \frac{Z_1}{\Lambda}
 - Z_2 \log \frac{Z_2}{\Lambda} + Z_3 \log \frac{Z_3}{\Lambda} \right)
\nonumber\\
a_{D2} &=&
-\frac{i}{\pi} \left(- Z_1 \log \frac{Z_1}{\Lambda}
 + 2 Z_2 \log \frac{Z_2}{\Lambda} + Z_3 \log \frac{Z_3}{\Lambda} \right)
\eeqa
yields that
\beqa
\Lambda = e^{- \frac{\langle S \rangle \pi}{6}}
\eeqa
(\ref{higgsad}), on the other hand, reproduces, up to an overall
minus sign,
equation (3.13) of Klemm et al \cite{KLT}.
The Higgs fields $a_1$ and $a_2$
transform as $a_1 \rightarrow a_2 - a_1, a_2 \rightarrow a_2$
under $\delta T \rightarrow - \rho^2
\delta U, \delta U \rightarrow - \rho^{-2} \delta T$.
It follows that the Higgs section transforms perturbatively as
\beqa
\left( \begin{array}{c}
a \\  a_D\\
\end{array} \right) \rightarrow {\tilde \Gamma}_{\infty}^{Higgs,w_1'}
\left( \begin{array}{c}
a \\  a_D\\
\end{array} \right) \;\;,\;\;
{\tilde \Gamma}_{\infty}^{Higgs,w_1'}
=
\left( \begin{array}{cccc}
-1&1&0&0  \\
0&1&0&0 \\
4&-2&-1&0\\
-2&1&1&1\\
\end{array} \right)
\eeqa
which reproduces equation (3.20) of Klemm et al \cite{KLT}.
Note that in Klemm et al \cite{KLT} one loops around singular points
in the opposite way we do.  Since the function we chose, equation
(\ref{fsu3}), has an opposite overall sign as compared
to their function (3.16), it follows that our and their
perturbative
monodromies should coincide, as they indeed do.

Finally note that, although $\Gamma_{\infty}^{w_2} =\Gamma_{\infty}^{w_1'}$
in the local case, the truncated monodromies
${\tilde \Gamma}_{\infty}^{Higgs,w_2}$ and
${\tilde \Gamma}_{\infty}^{Higgs,w_1'}$
are very different from each other.  This is due to the fact that the
associated 1-loop $f$ functions are very different and that the dilaton has
been frozen to $\langle S \rangle = \infty$ in the rigid case.

\subsection{The second Weyl twist $w_2'$ and the
third Weyl twist $w_0'$ of $SU(3)$}

Under the second Weyl twist $w_2'$
of $SU(3)$, the moduli $T$ and $U$ transform
as $T \rightarrow U + i, U \rightarrow T - i$.
Taking as the 1-loop corrected function $f(T,U)$ the one given in
(\ref{fsu3}), it can be checked using (\ref{spert}) that
$S \rightarrow S+i$.  Then, indeed,
the 1-loop corrected K\"{a}hler
potential $ K = - \log (Y_{tree} + Y_{pert}) $ is invariant under $w_2'$.
The resulting perturbative monodromy $\Gamma_{\infty}^{w_2'}$
is then given by
\beqa
\Gamma_{\infty}^{w_2'}=
\left( \begin{array}{cc}
U_{w_2'} & 0 \\ U^*_{w_2'}  \Lambda_{w_2'} & U^*_{w_2'}\\
\end{array} \right)
\eeqa
where $U_{w_2'}$ is given in (\ref{umatrices}) and where
\beqa
\Lambda_{w_2'} = \left( \begin{array}{cccc}
-2 & -1 & -2 & 2  \\ -1 & 0 & 0 & 0 \\
-2 & 0 & -2 & 1  \\ 2  & 0 &1 & -2 \\
\end{array} \right)
\eeqa

The perturbative monodromy $\Gamma_{\infty}^{w_0'}$
associated with the third Weyl twist is obtained from
$\Gamma_{\infty}^{w_1'}$ by conjugation as
\beqa
\Gamma_{\infty}^{w_0'} =
\left( \Gamma_{\infty}^{w_2'} \right)^{-1}
\Gamma_{\infty}^{w_1'}
\Gamma_{\infty}^{w_2'}=
\left( \begin{array}{cc}
U_{w_0'} & 0 \\ U^*_{w_0'}  \Lambda_{w_0'} & U^*_{w_0'}\\
\end{array} \right)
\eeqa
where $U_{w_0'}$ is given in (\ref{umatrices}) and where
\beqa
\Lambda_{w_0'} = \left( \begin{array}{cccc}
0&-3&-2&2\\
-3&0&-2&2\\
-2&-2&-4&3\\
2&2&3&-4
\end{array} \right)
\eeqa

Truncation to the rigid case is again achieved by freezing in
both the graviphoton and the dilaton, i.e.
$\langle P^0 \rangle = 1, \langle S \rangle = \infty$.
Due to the choice (\ref{fsu3})
of the 1-loop correction $f(T,U)$,
the resulting rigid monodromy matrices for the second and
the third Weyl twists
are again the ones given in
equation (3.20) of \cite{KLT}.

\subsection{Summary}

In summary, the complete
semiclassical monodromy is given by the product of the
four Weyl-reflection monodromies
times the monodromy matrix eq. (\ref{sdualpert})
which corresponds to
the discrete shifts in the
dilaton field.  In the following we will show how the four
perturbative monodromies
associated with the enhancement of gauge symmetries
are to be decomposed into non-perturbative monodromies due
to monopoles and
dyons becoming massless
at
points in the interior of moduli space.

\section{Non perturbative monodromies}

\setcounter{equation}{0}

\subsection{General remarks}

In order to obtain some information about non-perturbative
monodromies in $N=2$ heterotic string compactifications, we will follow
Seiberg/Witten's strategy in the rigid case \cite{SW1}
and try to decompose the
perturbative monodromy
matrices $\Gamma_{\infty}$ into
$\Gamma_{\infty} = \Gamma_M \Gamma_D$ with $\Gamma_M$ ($\Gamma_D$)
possessing monopole like (dyonic) fixed points. Thus each semi-classical
singular line will split into two non-perturbative singular lines where
magnetic monopoles or dyons respectively become massless. In doing so
we will
work in the limit of large dilaton field $S$ assuming that in this limit
the non-perturbative dynamics is dominated by the Yang-Mills gauge
forces. Nevertheless, the monodromy matrices we will obtain are not an
approximation in any sense, since the monodromy matrices are
of course field independent. They are just part of the full quantum
monodromy of the four-dimensional heterotic string.

Let us now precisely list the
assumptions we will impose when performing
the split of any of the semiclassical monodromies into the non-perturbative
ones:
\begin{enumerate}
\item
$\Gamma_{\infty}$ must be decomposed into precisely two factors.
\be
\Gamma_{\infty} = \Gamma_{M} \Gamma_{D}
\eq
\item
$\Gamma_{M}$ and therefore $\Gamma_{D}$ must be symplectic.
\item
$\Gamma_{M}$ must have a monopole like fixed point.
For the case of $w_1$, for instance, it must be of the form
\be
\left (N,-M \right) =
\left( 0,0,N^2, -N^2,0,0,0,0 \right)
\eq
\item
$\Gamma_{D}$ must have a dyonic fixed point.  For
the case of $w_1$, for instance, it must be of the form
\be
\left( N, -M \right) =
\left( 0,0,N^2, -N^2,0,0,-M_2, M_2 \right)
\eq
where $N^2$ and $M_2$ are proportional.
\item
$\Gamma_{M}$ and $\Gamma_{D}$ should be conjugated, that is, they must
be related by a change of basis, as it is the case in the rigid theory.
\item
The limit of large $S$ should be respected. This means that $S$
should only transform into a function of $T$ and $U$ (for at least
one of the four $SU(2)$ lines, as will be discussed in
the following).

\end{enumerate}

In the following we will show that under these assumptions the splitting
can be performed in a consistent way.  We will discuss the non perturbative
monodromies for the $SU(2)_{(1)}$ case in big detail.
Unlike
the rigid case, however, where
the decomposition of the perturbative monodromy into
a monopole like monodromy and a dyonic monodromy is unique (up to
conjugation), it will turn out that there are several distinct decompositions,
depending on four (discrete) parameters.
Only a subset
of these distinct decompositions should be,
however, the physically correct one.
One way of deciding which one is the physically correct one is to demand
that, when truncating this decomposition to the rigid case, one
recovers the rigid non perturbative monodromies of Seiberg/Witten.
This, however, requires one to have a reasonable prescription of
taking the flat limit, and one such prescription was given in section
(\ref{secseiwit}).

The non-perturbative part $f^{\rm NP}$ of the prepotential will depend on the
$S$-field.  We will make the following ansatz for the prepotential
\beqa
F =i \frac{X^1 X^2 X^3}{X^0} + (X^0)^2 \left(f(T,U) + f^{\rm NP} (S,T,U)
\right)
\eeqa
Then the non-perturbative period vector
$\Omega^T = (P, iQ)^T$
takes the form
\beqa
\Omega^T &=&
(1, TU - f^{\rm NP}_{S}
, iT, iU, iSTU + 2i (f + f^{\rm NP})
 - iT (f_T + f^{\rm NP}_{T})
- iU (f_U +  f^{\rm NP}_{U})  \nonumber\\
&-& iS  f^{\rm NP}_{S}
, iS,
- SU + f_T  +   f^{\rm NP}_{T}, -ST +
f_U +   f^{\rm NP}_{U})
\eeqa
This leads to the following non-perturbative mass formula for the
BPS states
\beqa
{\cal M} &=& M_I P^I + i N^I Q_I=
M_0 + M_1 (TU -  f^{\rm NP}_{S}) + i M_2 T + i M_3 U
+ i N^0 (STU \nonumber\\
&+& 2 (f + f^{\rm NP})
 - T (f_T + f^{\rm NP}_{T})
- U (f_U +  f^{\rm NP}_{U})  -S  f^{\rm NP}_{S})
+ i N^1 S \nonumber\\
&+& i N^2 ( i SU - i f_T  -i  f^{\rm NP}_{T} )
+ i N^3 ( i ST -i
f_U -i   f^{\rm NP}_{U})
\label{bpsmass}
\eeqa
Then we see that all states with $M_1\neq 0$ or $N^I\neq 0$ undergo
a non-perturbative mass shift. In the following we will use this
formula to determine (as a function of $f_{\rm NP}$ and its
derivatives) the singular loci where monopoles or dyons
become massless.  This will, for concreteness,
be done for the case of $SU(2)_{(1)}$.

\subsection{Non perturbative monodromies for $SU(2)_{(1)}$
\label{su21monopol}}

In order
to find a decomposition of
$\Gamma_{\infty}^{w_1}$, $\Gamma_{\infty}^{w_1}=
\Gamma_{M}^{w_1} \Gamma_{D}^{w_1}$,
we will now make the following ansatz:
$\Gamma_{\infty}^{w_1}$ has a peculiar block structure in that
the indices $j=0,1$ of the section $(P_j, i Q_j)$ are never
mixed with the indices $j=2,3$.
We will assume that $\Gamma_{M}^{w_1}$ and $\Gamma_{D}^{w_1}$ also have
this structure. This implies that the problem can be reduced
to two problems for 4 $\times$ 4 matrices.
Furthermore, we
will take $\Gamma_{M}^{w_1}$ to be the identity matrix
on its diagonal.
The existence of a basis
where the non--perturbative monodromies have this special form
will be aposteriori justified by the fact that it leads to
a consistent
truncation to the rigid case.

Then, let us first consider the submatrix of
$\Gamma_{\infty}^{w_1}$ which acts on $(P^2,P^3,iQ_2,iQ_3)^T$.  We
will show that its decomposition into non-perturbative pieces
is almost unique. More precisely, there will be a one
parameter family of decompositions, as follows.
The submatrix of
$\Gamma_{\infty}^{w_1}$ acting on $(P^2,P^3,iQ_2,iQ_3)^T$
is given by
\be
\Gamma_{\infty,23} =
\left [\begin {array}{cccc} 0&1&0&0\\\noalign{\medskip}1&0&0&0
\\\noalign{\medskip}1&-2&0&1\\\noalign{\medskip}-2&1&1&0\end {array}
\right ]
\eq
It will be decomposed into
$\Gamma_{\infty,23}=\Gamma_{M,23}
\Gamma_{D,23}$.  As stated above, we will make the following
ansatz for the monopole monodromy matrix $\Gamma_{M,23}$
\be
\Gamma_{M,23} =
\left [\begin {array}{cccc} 1&0&a&b\\\noalign{\medskip}0&1&c&d
\\\noalign{\medskip}p&q&1&0\\\noalign{\medskip}r&s&0&1\end {array}
\right ]
\eq
The existence of an eigenvector of the form
$(1,-1,0,0)$ implies that $p=q, r=s$, whereas symplecticity implies
$r=p, a=-b=-c=d$. Thus
\be
\Gamma_{M,23} =
\left [\begin {array}{cccc} 1&0&a&-a\\\noalign{\medskip}0&1&-a&a
\\\noalign{\medskip}p&p&1&0\\\noalign{\medskip}p&p&0&1\end {array}
\right ]
\eq
Computing the eigenvectors we find that the monopole fixed point
is unique (though the eigenvalue 1 has multiplicity 4).
Thus, $\Gamma_{M,23}$ appears to be reasonable. Computing $\Gamma_{D,23}$
we find
\be
\Gamma_{D,23} =
\left [\begin {array}{cccc} -3\,a&1+3\,a&a&-a\\\noalign{\medskip}1+3\,a&
-3\,a&-a&a\\\noalign{\medskip}1-p&-2-p&0&1\\\noalign{\medskip}-2-p
&1-p&1&0\end {array}\right ]
\eq
Requiring the existence of a dyonic fixed point
of $\Gamma_{D,23}$ fixes $a=-\frac{2}{3}$.
Moreover one automatically gets that $-M_2 =  \frac{3}{2} N^2$.
Hence
\be
\Gamma_{M,23} =
\left [\begin {array}{cccc} 1&0&-2/3&2/3\\\noalign{\medskip}0&1&2/3&
-2/3\\\noalign{\medskip}p&p&1&0\\\noalign{\medskip}p&p&0&1
\end {array}\right ],\;\;\;
\Gamma_{D,23} =
\left [\begin {array}{cccc} 2&-1&-2/3&2/3\\\noalign{\medskip}-1&2&2/3&
-2/3\\\noalign{\medskip}1-p&-2-p&0&1\\\noalign{\medskip}-2-p&1-p&1&
0\end {array}\right ]
\label{sub23}
\eq
For $p\not=0$ these matrices are conjugated, because they have the same
Jordan normal form.
This is, however, not the case if $p=0$. Naively one might have
expected this to be the natural choice because it makes
$\Gamma_{M,23}$ block triangular. But in the case
of $p=0$, $\Gamma_{M,23}$
has an additional eigenvector, whereas $\Gamma_{D,23}$ doesn't have one, and
hence
the matrices are not conjugated.

Next, consider the submatrix of
$\Gamma_{\infty}^{w_1}$ which acts on $(P^0,P^1,iQ_0,iQ_1)^T$.
Its symplectic decomposition
is less constrained.
Since we are in the perturbative
regime with respect to $S$, namely at $S=\infty$, we are not
looking for non--perturbative effects in the graviton/dilaton
sector, but only for non--perturbative effects in the gauge sector. Thus, the
decomposition of $\Gamma_{\infty,01}$
 should be
of the perturbative type.

This, on the other hand, gives a
three parameter family of decompositions of the perturbative
monodromy $\Gamma_{\infty,01}$, namely
\be
\Gamma_{\infty,01} =
\left [\begin {array}{cccc} 1&0&0&0\\\noalign{\medskip}0&1&0&0
\\\noalign{\medskip}0&-1&1&0\\\noalign{\medskip}-1&0&0&1\end {array}
\right ] =
\left [\begin {array}{cccc} 1&0&0&0\\\noalign{\medskip}0&1&0&0
\\\noalign{\medskip}x&y&1&0\\\noalign{\medskip}y&v&0&1\end {array}
\right ]
\cdot
\left [\begin {array}{cccc} 1&0&0&0\\\noalign{\medskip}0&1&0&0
\\\noalign{\medskip}-x&-y-1&1&0\\\noalign{\medskip}-y-1&-v&0&1
\end {array}\right ]
\nonumber\\
\eq
where both parts have no fixed point. They are
conjugated to each other, because
they have the same Jordan normal form.

Putting all these things together yields the following 8$\times$8
non--perturbative
monodromy matrices that consistently describe the splitting of
the $T=U$ line
\be
\Gamma_{M}^{w_1} =
\left [\begin {array}{cccccccc} 1&0&0&0&0&0&0&0\\\noalign{\medskip}0&
1&0&0&0&0&0&0\\\noalign{\medskip}0&0&1&0&0&0&-2/3&2/3
\\\noalign{\medskip}0&0&0&1&0&0&2/3&-2/3\\\noalign{\medskip}x&y&0&0&
1&0&0&0\\\noalign{\medskip}y&v&0&0&0&1&0&0\\\noalign{\medskip}0&0&p&p
&0&0&1&0\\\noalign{\medskip}0&0&p&p&0&0&0&1\end {array}\right ]
\label{monopolew1}
\eq
\be
\Gamma_D^{w_1} =
\left [\begin {array}{cccccccc} 1&0&0&0&0&0&0&0\\\noalign{\medskip}0&
1&0&0&0&0&0&0\\\noalign{\medskip}0&0&2&-1&0&0&-2/3&2/3
\\\noalign{\medskip}0&0&-1&2&0&0&2/3&-2/3\\\noalign{\medskip}-x&-y-1&0
&0&1&0&0&0\\\noalign{\medskip}-y-1&-v&0&0&0&1&0&0
\\\noalign{\medskip}0&0&1-p&-2-p&0&0&0&1\\\noalign{\medskip}0&0&-2-p
&1-p&0&0&1&0\end {array}\right ]
\label{dyonew1}
\eq

The associated fixed points have the form
\be
\left(N,-M \right) =
\left( 0 , 0 , N^2, -N^2, 0, 0, 0, 0 \right)
\eq
for the monopole and
\be
\left(N, -M \right) =
\left(0, 0,  N^2, -N^2, 0, 0, \frac{3}{2}
N^2, -\frac {3}{2} N^2 \right)
\eq
for the dyon.

\subsection{Truncating the $SU(2)_{(1)}$ monopole monodromy
to the rigid case}

The monopole monodromy matrix for the first $SU(2)$, given in
equation (\ref{monopolew1}),
depends on 4 undetermined parameters, namely
$x,v,y$ and $p\neq 0$.
Note that
demanding the monopole monodromy matrix to be conjugated to the dyonic
monodromy matrix led to the requirement $p \neq 0$.

On the other hand, it follows from (\ref{monopolew1}) that
\beqa S \rightarrow S - i \left( y + v (TU - f^{\rm NP}_S) \right)
\label{dilmonow1}
\eeqa

Consider now the $4 \times 4$ monopole subblock given in (\ref{sub23})
\beqa
\Gamma_{M23}^{w_1}=
\left [\begin {array}{cccc}
1&0&-2 \alpha &2 \alpha
\\\noalign{\medskip}0&1&2 \alpha &-2 \alpha
\\\noalign{\medskip}
p&p&1&0\\\noalign{\medskip}p&p&0&1\end {array}\right ] \;\;\;,\;\;\;
\alpha = \frac{1}{3} \;\;\;,\;\;\; p \neq 0
\eeqa
Rotating it into the Higgs basis gives that
\beqa
\Gamma_{M}^{Higgs,w_1}= M
\Gamma_{M23}^{w_1} M^{-1} =
\left [\begin {array}{cccc}
1&0&- 4\alpha \gamma^2&0
\\\noalign{\medskip}0&1&0&0
\\\noalign{\medskip}
0&0&1&0\\\noalign{\medskip}0&\frac{2p}{\gamma^2}&0&1\end {array}\right ]
\;\;\;,\;\;\; \alpha=\frac{1}{3} \;\;\;,\;\;\; p \neq 0
\label{locmonohiggs}
\eeqa
where $M$ is given in equation (\ref{higgsbasis}).
In the rigid case, on the other hand, one expects to find
for the rigid monopole monodromy matrix in the Higgs basis that
\beqa
\tilde{\Gamma}_{M}^{Higgs,w_1}=
\left [\begin {array}{cccc}
1&0&- 4\tilde{\alpha}\gamma^2&0
\\\noalign{\medskip}0&1&0&0
\\\noalign{\medskip}
0&0&1&0\\\noalign{\medskip}0&\frac{2\tilde{p}}{\gamma^2}
&0&1\end {array}\right ]
\;\;\;,\;\;\; \tilde{\alpha}=\frac{1}{4} \;\;\;,\;\;\; \tilde{p} =0
\label{rigidmono}
\eeqa
The first and third lines of (\ref{rigidmono}) are,
for $\tilde{\alpha}=\frac{1}{4}$, nothing but the monodromy matrix
for one $SU(2)$ monopole ($\gamma^2 = 2$ in the conventions of
Seiberg/Witten \cite{SW1}
, and $\gamma^2 = 1 $ in the conventions of Klemm et al \cite{KLT}).

Thus, truncating the monopole monodromy matrix (\ref{monopolew1})
to the rigid case appears to produce jumps in the parameters $p \rightarrow
\tilde{p}=0$ and $\alpha \rightarrow \tilde{\alpha}$ as given above.
In the following we would like to present a field theoretical explanation
for the jumps occuring in the parameters $p$ and $\alpha$ when
taking the rigid limit.

In the perturbative regime, that is at energies $E^2$ satisfying
$|\delta T + \delta U|^2 \gg
|\delta T-\delta U|^2 \gg E^2 \gg \Lambda^2$, we saw in subsection
(\ref{lamdil}) that $a_{D1}$ and $a_{D2}$ were given by
\beqa
a_{D1} &=& - \frac{i}{\gamma^2}  S  a_1  - \frac{4i}{\pi\gamma^2} a_1 \log
\left( \frac{\sqrt{2}}{\gamma} a_1\right) - \frac{2 i }{\pi\gamma^2} a_1 -
\frac{2}{\gamma^2} a_1
\nonumber\\
a_{D2} &=& \frac{i}{\gamma^2} S   a_2
\eeqa
Note that $a_{D2}$ didn't get any 1-loop correction in this regime.
On the other hand, as $E^2 \rightarrow \Lambda^2$, non-perturbative
corrections become important.  For $a_{D2}$ one expects these
non-perturbative corrections to be given by \cite{SW1}
\beqa
a_{D2} &=& \frac{i}{\gamma^2}  S   a_2 + \sum_{k\ge1} {\cal F}_k \left(
\frac{\Lambda}{a_2} \right)^{4k} a_2^2
\eeqa
However, since $|a_2 \propto \delta T + \delta U| \gg \Lambda$, it follows
that the non-perturbative corrections to $a_{D2}$ can here also be ignored,
that is $a_{D2} = \frac{i}{\gamma^2}  S
a_2$ in the regime under consideration.
For $a_{D1}$, on the other hand, the non-perturbative corrections
become important when $E^2 \rightarrow \Lambda^2$.

Now, under the monopole monodromy (\ref{monopolew1}) the dilaton shifts
as in (\ref{dilmonow1}), whereas $a_2 \rightarrow a_2$ as can be seen from
(\ref{locmonohiggs}).  Then it follows that
\beqa
a_{D2} &=& \frac{i}{\gamma^2}  S   a_2 \rightarrow
\frac{i}{\gamma^2}  \left( S - i \left[y + v (TU - f^{\rm NP}_S)
\right] \right) a_2 \nonumber\\
&=&
 a_{D2}
+ \frac{1}{\gamma^2}\left[y + v (TU - f^{\rm NP}_S)
\right]  a_2
\eeqa
Comparing with (\ref{locmonohiggs})
 shows that $v=0, 2p = y$ for
consistency.  Next, consider taking the rigid limit by freezing in the
dilaton to $\langle S \rangle$.  Then, under $a_2 \rightarrow  a_2$
it follows that
\beqa
a_{D2} = \frac{i}{\gamma^2} \langle  S \rangle  a_2 \rightarrow
\frac{i}{\gamma^2}  \langle S \rangle a_2 =
 a_{D2}
\eeqa
Thus, due to the freezing in of the dilaton field, one finds that $p\neq 0
\rightarrow \tilde{p} = 0$!

Next, consider the dynamically generated scale $\Lambda$ which, in the local
case, is given by
\beqa
\Lambda = e^{- \frac{\pi}{4}  S
- \log \frac{\sqrt{2}}{\gamma} + \frac{i \pi}{2}}
\eeqa
Under (\ref{dilmonow1}), it follows that $\log \Lambda$
transforms into
\beqa
\log \Lambda \rightarrow \log \Lambda + \frac{i \pi}{4}
\left( y + v (TU - f^{\rm NP}_S) \right)
\eeqa
which for $v=0$ turns into
\beqa
\log \Lambda \rightarrow \log \Lambda + \frac{i \pi}{4} y
\label{tranlammono}
\eeqa

In the rigid case, as $E^2 \rightarrow \Lambda^2$,
$a_1$ was determined by Seiberg/Witten to be given by
\beqa
a_1 &=& constant - \frac{2 i \tilde{\alpha}\gamma^2 }{\pi}
 a_{D1} \log \frac{a_{D1}}{\Lambda} \;\;\;,\;\;\;
\tilde{\alpha}=\frac{1}{4} \nonumber\\
a_{D1} &=& c_0 (u - \Lambda^2)
\eeqa
Indeed, as $u - \Lambda^2 \rightarrow e^{-2i\pi}(u - \Lambda ^2)$, $a_{D1}
\rightarrow e^{-2i \pi} a_{D1}$, it follows that
\beqa
a_1 &\rightarrow& a_1 - 4 \tilde{\alpha}\gamma^2 \, a_{D1} \nonumber\\
a_{D1} & \rightarrow& a_{D1}
\eeqa
which is consistent with (\ref{rigidmono}).
The 1-loop contribution to $a_1$ can also be understood as arising from
a Feynman graph in the dual theory with 2 external magnetic photon lines
and a light monopole hypermultiplet of mass $m \propto a_{D1}$
running in the loop.  The 1-loop
beta function coefficient is proportional to
$\tilde{\alpha}$.

In the local case, on the other hand, nothing changes in the computation of
this magnetic Feynman graph.  Thus, in the local case one
has again that
\beqa
a_1 &=& constant - \frac{2 i \tilde{\alpha}\gamma^2 }{\pi}
 a_{D1} \log \frac{a_{D1}}{\Lambda} \;\;\;,\;\;\;
\tilde{\alpha}=\frac{1}{4}
\eeqa
A crucial difference, however, arises in that the dynamically generated
scale $\Lambda$ now transforms as well under modular transformations,
namely as given in (\ref{tranlammono}).
Then, it follows that
\beqa
a_1 &\rightarrow& a_1 - \frac{2i \tilde{\alpha}\gamma^2}{\pi}
(-2i\pi - \frac{i\pi}{4}y)
 a_{D1} = a_1 - 4 \alpha\gamma^2 \, a_{D1}
  \nonumber\\
a_{D1} & \rightarrow& a_{D1}
\eeqa
where $ \alpha = \tilde{\alpha} (1+\frac{y}{8})$.
Thus, one sees that the jump in $\alpha \rightarrow \tilde{\alpha}$
when taking the rigid limit is a direct consequence of the
freezing in of the dilaton.  Finally, with $\tilde{\alpha}= \frac{1}{4}$
and $\alpha=\frac{1}{3}$ it follows that
$y = \frac{8}{3}$ and that $p = \frac{4}{3}$.

Thus, we have given a field theoretical explanation for the jumping
occuring in certain parameters when taking the rigid limit.  As a
bonus we have also been able to determine the value of the
parameters $v,y$ and $p$.  Moreover, one can show that, in order
to decouple the four $U(1)$'s at the non-perturbative level, one
has to have $x=v$ and consequently $x=0$.  It is, indeed, reasonable
to have $x=0$ because then the nonperturbative monodromy matrices
(\ref{monopolew1}) and (\ref{dyonew1}) become symmetric with
respect to $T$ and $U$.
Note that $v=0$
ensures that $S \rightarrow S - i y$ under the $SU(2)_{(1)}$
monopole monodromy.

\subsection{Singular loci for $SU(2)_{(1)}$}

Let us consider the Weyl twist $w_1$ in the first $SU(2)$.
The associated monopole eigenvector has non vanishing
quantum numbers $N^3=-N^2$.  Then, it follows from (\ref{bpsmass})
that its mass vanishes for $Q_2=Q_3$, which gives that
\beqa
i S(T-U) - i (f_T - f_U) - i ( f^{\rm NP}_T -  f^{\rm NP}_U ) =0
\label{locusmw1}
\eeqa
Under the monopole monodromy (\ref{monopolew1}), it follows
that
\beqa
T \rightarrow T - \frac{2}{3} ( Q_2 - Q_3) \nonumber\\
U \rightarrow U +\frac{2}{3} ( Q_2 - Q_3)
\eeqa
Then, on the locus of vanishing monopole masses (\ref{locusmw1}), one
has that $T\rightarrow T, U \rightarrow U$.

The associated dyon eigenvector, on the other hand, has
non vanishing quantum numbers $M_3 = - M_2 = \frac{3}{2} N^2,
N^3 = - N^2$.  Then, it follows from (\ref{bpsmass}) that
its mass vanishes for
\beqa
T - U  = \frac{2}{3} (Q_2 - Q_3)
\label{locusdw1}
\eeqa
Under the dyon monodromy (\ref{dyonew1}), it follows that
\beqa
T \rightarrow - U
 + 2 T -\frac{2}{3} (Q_2 - Q_3) \nonumber\\
U \rightarrow -T + 2 U + \frac{2}{3} (Q_2 - Q_3)
\eeqa
On the locus of vanishing dyon masses (\ref{locusdw1})
one then has again that $T \rightarrow T, U \rightarrow U$.

Similar considerations can be made for any of the other 3
$SU(2)$ lines.

\subsection{Non perturbative decomposition of the other 3 $SU(2)$ lines}

As discussed in section \ref{classres}, the perturbative monodromy matrices
associated with the 4 $SU(2)$ lines are conjugated to each other by
the generators $\sigma, g_1$ and $g_2$.  Then, it follows
that the non-perturbative
decomposition of any of the perturbative monodromies associated
with $w_2$, $w_1'$ and $w_2'$ is conjugated to the non-perturbative
decomposition given above for $\Gamma_{\infty}^{w_1}$.
For concreteness, we will below show how the non-perturbative
monodromies of $SU(2)_{(2)}$ can be obtained from the ones
of $SU(2)_{(1)}$ by conjugation with the generator $g_1$.
We will find one additional monopole and one additional dyon
eigenvector for $SU(2)_{(2)}$.  An analogous decomposition
of the remaining perturbative matrices associated with
$w_2'$ and $w_0'$ leads to 1 additional monopole and to 3
additional dyons.  Thus,
similarly to what one has in the rigid case, one finds 2
monopoles and 2 dyons for the case of
$SU(2)_{(1)} \times SU(2)_{(2)}$, whereas for the $SU(3)$ case
one finds 2 monopole and 4 dyon eigenvectors, which
are conjugated to each other \cite{KLTY,KLT}.

The explicit matrix representation
of the generator
$g_1$ is
\be
\Gamma^{(g_1)} =
\left [\begin {array}{cccccccc} 0&0&1&0&0&0&0&0\\\noalign{\medskip}0&0
&0&1&0&0&0&0\\\noalign{\medskip}-1&0&0&0&0&0&0&0\\\noalign{\medskip}0&
-1&0&0&0&0&0&0\\\noalign{\medskip}0&0&0&0&0&0&1&0\\\noalign{\medskip}0
&0&0&0&0&0&0&1\\\noalign{\medskip}0&0&0&0&-1&0&0&0\\\noalign{\medskip}0
&0&0&0&0&-1&0&0\end {array}\right ]
\eq
where $\left( \Gamma^{(g_1)} \right)^2 = - I$.
The perturbative and non-perturbative
monodromies for $SU(2)_{(2)}$ are obtained from the
monodromies of $SU(2)_{(1)}$ by conjugation with $\Gamma^{(g_1)}$,
$\Gamma^{(w_2)}_{\infty,M,D} =
(\Gamma^{(g_1)})^{-1} \Gamma^{(w_1)}_{\infty,M,D} \Gamma^{(g_1)}$.
They are computed to be
\be
\Gamma^{(w_2)}_{\infty} =
\left [\begin {array}{cccccccc}
0&1&0&0&0&0&0&0\\\noalign{\medskip}1&0
&0&0&0&0&0&0\\\noalign{\medskip}0&0&1&0&0&0&0&0\\\noalign{\medskip}0&0
&0&1&0&0&0&0\\\noalign{\medskip}1&-2&0&0&0&1&0&0\\\noalign{\medskip}-2
&1&0&0&1&0&0&0\\\noalign{\medskip}0&0&0&-1&0&0&1&0\\\noalign{\medskip}0
&0&-1&0&0&0&0&1\end {array}\right ]
\eq
\be
\Gamma^{(w_2)}_{M} =
\left [\begin {array}{cccccccc} 1&0&0&0&-2/3&2/3&0&0
\\\noalign{\medskip}0&1&0&0&2/3&-2/3&0&0\\\noalign{\medskip}0&0&1&0&0
&0&0&0\\\noalign{\medskip}0&0&0&1&0&0&0&0\\\noalign{\medskip}p&p&0&0&1
&0&0&0\\\noalign{\medskip}p&p&0&0&0&1&0&0\\\noalign{\medskip}0&0&x&y
&0&0&1&0\\\noalign{\medskip}0&0&y&v&0&0&0&1\end {array}\right ]
\eq
\be
\Gamma^{(w_2)}_{D} =
\left [\begin {array}{cccccccc} 2&-1&0&0&-2/3&2/3&0&0
\\\noalign{\medskip}-1&2&0&0&2/3&-2/3&0&0\\\noalign{\medskip}0&0&1&0&0
&0&0&0\\\noalign{\medskip}0&0&0&1&0&0&0&0\\\noalign{\medskip}1-p&-2-p
&0&0&0&1&0&0\\\noalign{\medskip}-2-p&1-p&0&0&1&0&0&0
\\\noalign{\medskip}0&0&-x&-y-1&0&0&1&0\\\noalign{\medskip}0&0&-y-1&-
v&0&0&0&1\end {array}\right ]
\eq
First note that now $P_0$ transforms into some $Q_i$, and therefore
the constraint $S=\infty$ seems to be violated. However, since
now something non--trivial has to happen with the quantum
numbers $N_0, N_1$ which are related to the magnetic quantum
numbers of $SU(2)_{(2)}$, it is inevitable, that some non--vanishing
entries appear at that place. Moreover, the physics should be
the same as on the line $T=U$ because both sets of matrices
are conjugated by a perturbative monodromy transformation.

The associated fixed points have the expected form, namely
\be
\left(N,-M \right) =
\left( -N^2, N^2, 0, 0, 0, 0, 0, 0 \right)
\eq
for the monopole and
\be
\left(N, -M \right) =
\left( -N^2, N^2, 0, 0, -\frac{3}{2}
N^2, \frac {3}{2} N^2, 0, 0 \right)
\eq
for the dyon.

\section{Conclusions}

We have shown in the context of four-dimensional heterotic strings
that the semiclassical monodromies associated with lines of enhanced
gauge symmetries can be consistently split into pairs of non-perturbative
lines of massless monopoles and dyons. Furthermore, all monodromies
obtained in the string context allow for a consistent truncation to
the rigid monodromies of \cite{SW1,KLTY,KLT}.
It would be very interesting to compare the monodromies obtained
on the heterotic side with computations on the type II side of
monodromies in appropriately chosen Calabi-Yau spaces.

In this paper we have not addressed the splitting
of the semiclassical monodromy (\ref{sdualpert}), associated with
discrete shifts in the $S$ field, into non-perturbative
monodromies.  If indeed such a splitting occurs, then it
should be due to new gravitational stringy non-perturbative effects
occuring at finite $S$, i.e. $ S \approx 1$.

\section{Acknowledgement}

We would like to thank P. Candelas, G. Curio, X. de la Ossa,
E. Derrick, V. Kaplunovsky, W. Lerche,
J. Louis and S. Theisen for fruitful discussions.
One of us (D.L.) is grateful to the Aspen Center of Physics, where
part of this work was completed.  The work of G.L.C. is supported
by DFG.

\section{Note added}

After completion or our work non-perturbative monodromies were
computed \cite{KKLMV,AntoPartou} using the string-string duality
between the $N=2$ heterotic and type II strings.  Specifically, for
the rank two model with fields $S$ and $T$ it was shown that the
type II Calabi-Yau monodromies at the conifold points correspond
to the non-perturbative heterotic monodromies due
to massless monopoles and dyons.  The perturbative monodromy (in
this case $T \rightarrow \frac{1}{T}$) and its decomposition into
non-perturbative monopole and dyon monodromies, as computed from
the type II Calabi-Yau side, agree with our perturbative and
non-perturbative monodromies
after introducing a compensating
shift for the dilaton.  This corresponds to a different, but
equivalent freezing in of the dilaton.

\end{document}